\documentclass[traditabstract, a4paper]{aa}

\usepackage{amsmath}
\usepackage{txfonts}
\usepackage{natbib}
\usepackage{color}
\usepackage[normalem]{ulem} 
\usepackage{comment}
\usepackage{ifthen}
\usepackage{graphicx,subfigure}

\specialcomment{blockcomment}{ \textbf{TO BE DELETED:} \begingroup\textcolor{green}}{\endgroup}

\ifthenelse{\isundefined{\withnotes}}{
  \excludecomment{blockcomment}  
  \newcommand{\mycomment}[1]{}   
  \newcommand{\strike}[1]{}
  \newcommand{\mikes}[1]{}  
  \newcommand{\mike}[1]{#1}

}{
  \newcommand{\mycomment}[1]{\textbf{#1}} 
  \newcommand{\mike}[1]{\textbf{#1}} 

  \ifthenelse{\isundefined{\nostrikes}}{
  	\newcommand{\strike}[1]{\sout{#1}}
  	\newcommand{\mikes}[1]{\textcolor{red}{\sout{#1}}}
  }{
  	\newcommand{\strike}[1]{}
  	\newcommand{\mikes}[1]{}
  	\excludecomment{blockcomment}
  }
}

\begin{document}

\title{Faraday caustics}
\subtitle{Singularities in the Faraday spectrum and their utility as probes of magnetic field properties} 

\author{M.R.~Bell\inst{\ref{inst:mpa}} \and H.~Junklewitz\inst{\ref{inst:mpa}} \and T.A.~En\ss lin\inst{\ref{inst:mpa}}}

\institute{Max Planck Institute for Astrophysics, Karl-Schwarzschild-Str. 1, 85741 Garching, Germany \email{mrbell@mpa-garching.mpg.de}\label{inst:mpa}} 

\date{Received \textbf{???} / Accepted \textbf{???}} 

\abstract{We describe singularities in the distribution of polarized intensity as a function of Faraday depth (i.e. the Faraday spectrum) caused by line-of-sight (LOS) magnetic field reversals. We call these features Faraday caustics because of their similarity to optical caustics.  They appear as sharply peaked and asymmetric profiles in the Faraday spectrum, that have a tail that extends to one side.  The direction in which the tail extends depends on the way in which the LOS magnetic field reversal occurs (either changing from oncoming to retreating or vice versa). We describe how Faraday caustics will form three-dimensional surfaces that relate to boundaries between regions where the LOS magnetic field has opposite polarity.  We present examples from simulations of the predicted polarized synchrotron emission from the Milky Way. We derive either the probability or luminosity distribution of Faraday caustics produced in a Gaussian magnetic field distribution as a function of their strength, $\mathcal{F}$, and find that for strong Faraday caustics $P(\mathcal{F}) \propto \mathcal{F}^{-3}$.  If fully resolved, this distribution is also shown to depend on the Taylor microscale, which relates to the largest scale over which dissipation is important in a turbulent flow.} 


\maketitle


\section{Introduction}\label{sec:intro}

The availability of broadband receivers in the next generation of radio telescopes, such as the Expanded Very Large Array (EVLA), the Square Kilometer Array (SKA), and its pathfinder arrays such as the Low-Frequency Array (LOFAR), has led to the development of a novel technique for imaging polarized radio emission known as rotation measure (RM) synthesis \citep{brentjens_faraday_2005}. Derived from a technique originally proposed by \citet{burn_depolarization_1966}, RM synthesis allows for the separation of polarized sources along the line of sight (LOS) by reconstructing the distribution of polarized emission as a function of Faraday depth (in the simplest case, Faraday depth is equivalent to RM). The technique provides several benefits: sensitivity is improved by combining measurements from hundreds or even thousands of channels over a broad frequency range, the rotation measure of point sources can be measured more accurately and unambiguously, and the intrinsic polarized emission from different sources along a line of sight can be studied independently.  Since the Faraday depth depends on the integrated LOS magnetic field, RM synthesis also promises to be an important tool for studying magnetism.

To date, there have been few applications of RM synthesis, although interest is rapidly increasing as new radio telescopes are being commissioned.  Its applications have included studies of the diffuse polarized emission in the Perseus field \citep{de_bruyn_diffuse_2005, brentjens_wide_2010} and that associated with Abell 2255 \citep{pizzo_deep_2010}, analysis of the polarized emission in nearby galaxies in the WSRT-SINGS survey \citep{heald_westerbork_2009}, and the detection of a shell of compressed magnetic field surrounding a local HI bubble \citep{wolleben_antisymmetry_2010}.  The RM synthesis technique will play a critical role in several upcoming polarization surveys, e.g. POSSUM \citep{gaensler_possum_2010}, GMIMS \citep{wolleben_gmims_2009}, and future surveys with LOFAR.  

The technique of RM synthesis builds on the work of \citet{burn_depolarization_1966} who showed that the complex polarized intensity, $P$, as a function of wavelength-squared ($\lambda^2$), is related to the intrinsic complex polarized intensity as a function of Faraday depth, $F$, by a Fourier transformation
\begin{equation}
	P(\lambda^2) = \int_{-\infty}^{\infty} F(\phi, \lambda^2) \text{e}^{2i\phi \lambda^2} d\phi.
\label{eq:fourier_relation}
\end{equation}
The polarized intensity is $P=Q+iU$, where $Q$ and $U$ are the Stokes parameters that describe the linear polarization state of the emission.  The coordinate $\phi$, known as the \emph{Faraday depth}, is a measure of the amount of rotation suffered by a linearly polarized wavefront as it passes through a magneto-ionic medium.  It is measured in rad m$^{-2}$ and is given by 
\begin{equation}
	\phi = a_0 \int _0^z n_e(z') B_{3}(z') dz'
	\label{eq:faraday_depth}
\end{equation}
where $n_e$ is the number density of thermal electrons, $z$ is distance along the LOS, and $B_{3}(z')$ is the component of the magnetic field along the LOS. The constant term is $a_0=e^3/(2 \pi m_e^2 c^4)=0.81$ cm$^3$ ($\mu$G pc)$^{-1}$, where $e$ and $m_e$ are the charge and mass of the electron, respectively.

The quantity $F(\phi, \lambda^2)$ is the intrinsic polarized intensity as a function of Faraday depth.  We refer to this quantity as the Faraday spectrum.  It is also commonly referred to as the Faraday dispersion function following \citet{burn_depolarization_1966}. Though $F$ generically depends on the wavelength, we will assume a separable form, $F(\phi, \lambda^2) = f(\phi) s(\lambda^2)$, allowing for the removal of $s(\lambda^2)$ from the integral in Eq. \eqref{eq:fourier_relation}. We will henceforth assume that $P(\lambda^2)$ can be normalized by the spectral dependence such that $P(\lambda^2) = \int F(\phi) \text{e}^{2i\phi \lambda^2} d\phi$. 

The Faraday spectrum is a three dimensional description of the polarized emission and may be reconstructed by inverting Eq. \eqref{eq:fourier_relation}.  In general, this is not possible because of our inability to completely sample $P$ for all values of $\lambda^2$, but this complication is resolved using the RM synthesis technique of \citet{brentjens_faraday_2005} where the inversion is treated in much the same way as it is in synthesis imaging. The incomplete sampling of $P(\lambda^2)$ is represented by introducing a sampling function.  The reconstructed Faraday spectrum is then a convolution of the true Faraday spectrum and the Fourier transform of the sampling function, known as the RM spread function (RMSF).  The RMSF is a point spread function in $\phi$, and is analogous to the ``dirty beam'' in synthesis imaging.

In this paper we are concerned with the properties of the Faraday spectrum of extended synchrotron emitting and Faraday rotating sources.  In particular, we describe singularities in the Faraday spectrum that we call \emph{Faraday caustics}.  These arise when the LOS magnetic field changes direction, causing a pile-up of emission at a single Faraday depth.  They are caustics, much like the well known optical phenomenon, in the sense that they are the result of the non-monotonic mapping of the polarized emissivity from physical depth space into Faraday depth space. Optical caustics, for instance those seen at the bottom of a swimming pool on a sunny day, are also caused by a non-monotonic mapping of the surface radiation field to its lensed image at the bottom.

Before we derive a mathematical description of Faraday caustics, it is helpful to develop a qualitative understanding of why they occur.  We consider a volume filled with magnetic fields, and both relativistic and thermal electrons.  This volume will produce synchrotron emission and rotate the plane of polarization of a propagating plane wave.  The polarized emission from each location along a LOS generally maps to a different location in Faraday depth space. Therefore we expect to find extended structures in RM synthesis observations of such a source.  When the LOS magnetic field component is changing sign, what happens near the point where the field is completely in the plane of the sky?  In this case, the LOS field is close to zero and the Faraday depth will change very little with physical depth because, according to Eq. \eqref{eq:faraday_depth}, the rate of change in Faraday depth with $z$ varies linearly with the LOS magnetic field component.  This means that all of the polarized emission originating from this region will pile up around a single Faraday depth value and we expect the resulting Faraday spectrum to be strongly peaked.

In this paper, we have two aims. First, we wish to inform observers about the fundamental properties of Faraday caustics. We expect them to be ubiquitous in the Faraday spectra of extended polarized sources, hence they should be taken into account when interpreting observational data.  We first describe the basic features of Faraday caustics using a simple description of a magnetic field reversal, and provide examples of how they appear in simulations of the Galactic synchrotron emission.  Second we investigate how observations of Faraday caustics can be exploited to uncover information about the structure and statistical properties of the underlying magnetic field.  We present a few ideas and examples.  This second intention is in line with previous work on the extraction of statistical information of magnetic fields from observables by \citet{spangler_turbulent-transport_1982, spangler_mag-turbulence_1983}, \citet{eilek_turbulence_1989a, eilek_turbulence_1989b}, \citet{ensslin_vogt_2003a}, \citet{kahniashvili_helicity_2006}, \citet{mizeva_polarization-stats_2007}, \citet{fletcher_canals_2007}, \citet{waelkens_mag-field-stats_2009}, \citet{junklewitz_magnetic_power_2010}, and \citet{oppermann_litmus_2010}.

In Sect. \ref{sec:basic_features}, we derive the properties of Faraday caustics by considering a simple magnetic field near the location where it is aligned in the plane of the sky. We examine the observational properties of caustics in Sect. \ref{sec:obs_examples}, presenting examples of simulated synchrotron emission from the plane of the Milky Way. The luminosity function of caustics generated in a Gaussian random magnetic field as a function of their strength is computed in Sect. \ref{sec:caustics_in_gauss_field}.  Finally, in Sect. \ref{sec:conclusions} we present our conclusions.

\section{Properties of Faraday caustics}\label{sec:basic_features}

Faraday caustics appear where the magnetic field lies entirely in the plane of the sky, i.e. where $B_z=0$, in a mixed synchrotron emitting and Faraday rotating medium.  To characterize the basic features of Faraday caustics, we assume the simplest magnetic field configuration possible and focus on the region immediately surrounding the location where the LOS component is zero. We label this point $z_0$, where $B_3(z_0) = 0$, and henceforth for simplicity we set $z_0 = 0$ (the observer is located at $z<0$).  Ultimately, we wish to solve for the Faraday spectrum, $F(\phi)$.

The Faraday spectrum is related to the intrinsic polarized intensity, $P(z)$, by    

\begin{equation}
	F(\phi) = \int_{-\infty}^\infty \delta(\phi - \phi(z)) P(z) dz.
\label{eq:pz_to_fphi_def}
\end{equation}
The polarized intensity per unit length along $z$ is 
\begin{align}
	P(z)dz & = p(\alpha) I(z) e^{2i\chi(z)} dz \notag \\
	& = a_s(\alpha, z) \left|B_\perp(z)\right|^{1+\alpha} e^{2i\chi(z)} dz.
\label{eq:pz_def}
\end{align}
To simplify the following expressions, we encapsulate all non-essential terms into $a_s$.  This parameter therefore contains all of the constant and frequency-dependent terms in the expression for the synchrotron emissivity, $I(z)$ (see e.g. \citealp{rl04}), as well as the fractional polarization of synchrotron emission, $p(\alpha)=(\alpha+1)/(\alpha+5/3)$.  It varies with $z$ because it also incorporates the dependence on the cosmic-ray electron density.  In general, this will be essential for understanding the detailed structure of a Faraday caustic profile, but in this illustrative case we assume that the cosmic-ray electron density is roughly constant in the region around $z_0=0$. The parameter $\alpha$ is the spectral index, where $I(\nu) \propto \nu^{-\alpha}$.  For convenience, we assume that $\alpha=1$ throughout the remainder of this paper. This assumption is approximately correct for many optically thin synchrotron sources and suits our purpose of providing a basic description.  The angle $\chi$ is the intrinsic position angle and $B_{\perp}$ is the magnetic field in the plane of the sky.  To compute $F$, we require a description of the $z$ dependence of both $B_3$ and $B_{\perp}$.

We restrict our attention to the region immediately surrounding $z_0=0$ and expand the LOS magnetic field around $z_0$ to first order with respect to $z$ as

\begin{equation}
	B_{3}(z) = B_{3}'(0)z + \ldots
\label{eq:bz}
\end{equation}
where the prime indicates differentiation in $z$.  We recall that $B_3(0)=0$, by definition. In this way, we capture the essential properties of a field reversal without having to specify a functional form for $B_3(z)$.  From Eq. \eqref{eq:faraday_depth}, we find that
\begin{equation}
	\phi(z) - \phi_0 = \frac{a_1 B'_3(0) z^2}{2},
\label{eq:phiz}
\end{equation}
where $a_1 = a_0 n_e$ and we assume that $n_e(z)$ is approximately constant over this limited range in $z$. Our illustrative example of course becomes inaccurate for distances over which $n_e$ starts to vary significantly.  Such a variation will not affect the basic properties of a caustic, so we ignore it for the moment.  

At this point, we can get an idea about the general form of $F(\phi)$.  The last expression tells us how the polarized emission originating at a given $z$ location maps to $\phi$-space; adopting the linear approximation for the $z$-component of the magnetic field given by Eq. \ref{eq:bz}, we can see that $\phi$ is quadratic in $z$. The value $\phi(z)=\phi_0$, which occurs when $z=z_0=0$, is either a maximum or minimum depending on the sign of $B'_3(0)$. Therefore, the Faraday spectrum will extend only to one side or the other of $\phi_0$. Each $\phi$-location on this side of $\phi_0$ will be mapped to the two physical locations given by the multivalued inverse, $z(\phi)$.  In addition, as $z \rightarrow 0$ the change in Faraday depth, $d\phi/dz$, becomes small and the emission over a potentially large range of $z$ values ``piles up'' over a small range of $\phi$ values.  As a result, we expect that $F(\phi)$ will sharply increase near $\phi_0$ and appear as an asymmetric ``spike''.  Smaller values of $\left|B'_3\right|$ lead to a flatter $\phi(z \sim 0)$, hence we then expect to find stronger spikes when the LOS magnetic field changes direction over larger distances.

\mike{Since we are primarily concerned with describing the effect of a LOS magnetic field reversal, we assume the simplest possible form of $B_\perp(z)$.  First, we assume that the orientation of $B_\perp$ is fixed over the limited range of $z$ that we are interested in and that it is aligned along the $x$-axis. Our magnetic field is then}

\begin{equation}
	\vec{B} = B_\perp(x,z) \: \vec{\hat{x}} + B_3'(0)z \: \vec{\hat{z}}.
	\label{eq:total_b_field}
\end{equation}  
\mike{To determine $B_\perp(x,z)$, we use the condition that the magnetic field is divergence-free, which leads to}
	
\begin{equation}
	B_\perp(x,z) = B_\perp(0,z) - B'_3(0) x.
	\label{eq:b_perp_linear}
\end{equation}
\mike{With no loss of generality, we choose the LOS in question to be along $x=0$.  Also, for simplicity, we assume that there is no variation in $B_\perp(0,z)$ over the small portion of the line of sight in which we are interested, i.e.}

\begin{equation}
	B_\perp(0,z) \approx B_\perp(0,0) = const.
	\label{eq:b_perp_const}
\end{equation}
With the magnetic fields specified, we can now calculate $F$.  From Eqs. \eqref{eq:pz_to_fphi_def}, \eqref{eq:pz_def}, \eqref{eq:phiz}, and \eqref{eq:b_perp_const}, we find that

\begin{align}
	F(\phi) = & \int dz \bigg[ \delta\left(\phi - \phi_0 - \frac{a_1 B'_3(0) z^2}{2}\right) \notag \\ 
	& a_s \left|B_\perp(0)\right|^{2} e^{2i\chi} \bigg].
\label{eq:pz_to_fphi}
\end{align}
We recall that we assumed that $B_\perp$ has a fixed orientation, which implies that $\chi$ is also fixed.

To perform the $z$ integration, we use the property of the delta function
\begin{equation}
	\delta\left(g(x)\right)=\sum_{\overline{x}:g(\overline{x})=0} \frac{\delta\left(x-\overline{x}\right)}{g'(\overline{x})},
\label{eq:delta_function_property}
\end{equation}
where $\overline{x}$ are the roots of $g(x)$, the sum is over all roots, and the prime indicates differentiation with respect to x.  The argument of the delta function has two roots, $\overline{z}=\pm\sqrt{2(\phi-\phi_0)/(a_1 B'_z(0))}$.  This means that the polarized emission from two values of $z$ will contribute to $F$ at a single value of $\phi$.  After performing the $z$ integration, we find that

\begin{equation}
	F(\phi) = \frac{D \: \Theta[(\phi - \phi_0)B_3'(0)]}{\sqrt{\phi-\phi_0}},
\label{eq:Fphi}
\end{equation}
where 

\begin{equation}
D = \frac{2 a_s \left| B_\perp(0)\right|^2}{\sqrt{a_1 B_3'(0)}}
\label{eq:CD_consts}
\end{equation}
and $\Theta$ is the Heaviside function.

\begin{figure}%
	\centering
	\includegraphics[width=8cm]{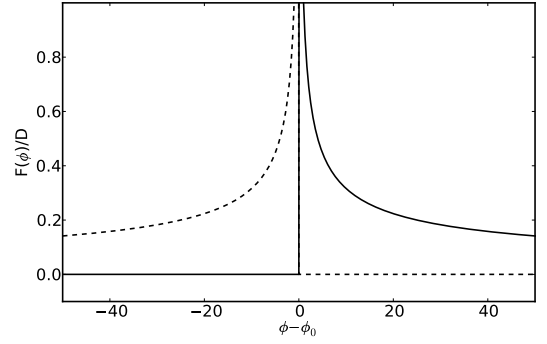}
	\caption{The normalized Faraday spectrum, $F(\phi-\phi_0)/D$, for a caustic.  The solid and dashed lines apply when $B_3'(0) > 0$ and $B_3'(0) < 0$, respectively.} 
	\label{fig:fphi}%
\end{figure}

The final form of $F$, as shown in Fig. \ref{fig:fphi}, is as we expect.  The spectrum diverges at $\phi=\phi_0$, and is asymmetric with a tail extending to one side or the other of $\phi_0$.  The direction of the tail is determined by the sign of $B_3'(0)$, as indicated by the Heaviside function in Eq. \ref{eq:Fphi}.

Since in many cases caustics will be unresolved in observations (see Sect. \ref{sec:obs_examples}), we should consider the integrated intensity over the range $\delta \phi$, which is taken to be approximately the FWHM of the point spread function in $\phi$ (often called the rotation measure spread function or RMSF).  The integrated intensity, $\mathcal{F}$, is

\begin{equation}
	\mathcal{F} = \left| \int_{\phi_0}^{\phi_0 + \delta \phi} d\phi \ F(\phi) \right|.
\label{eq:int_f_setup}
\end{equation} 
Together with Eqn. \ref{eq:Fphi}, this gives

\begin{equation}
	\mathcal{F} = \left|2 D \sqrt{\delta \phi} \right| =  4 a_s \left|B_{\perp}(0)\right|^2 \sqrt{\frac{\delta \phi}{a_1 B'_3(0)}}.
\label{eq:int_f_lowu}
\end{equation}
As expected, we find that small values of $B'_3$ lead to strong caustic features.  Large $B_{\perp}$ values will also result in large values of $\mathcal{F}$.  However, since the $B$-field strength is unlikely to deviate significantly from the average and the $1/\sqrt{B'_3}$ term can be arbitrarily large, we argue that strong caustics will mostly be created in regions where $B'_3$ is small.  A result of this effect is that large ``spikes'' in the Faraday spectrum can result from intrinsically weak emission regions in physical space.

We have only considered an illustrative case above, assuming constant thermal and relativistic electron fields, and the simplest possible magnetic field that includes a LOS reversal.  More complicated fields will not have a significant effect on the fundamental structure of Faraday caustics.  Regardless of the form of the LOS magnetic field reversal, a caustic will appear as a strongly peaked, asymmetric source in Faraday space.  The higher order terms in the expansion of $B_3$, and the variation in the particle and $B_\perp$ fields, will affect the specific shape of the ``tail.'' Equation \ref{eq:Fphi} is the asymptotic form of $F$ for a caustic when approaching $\phi_0$.  The range of $\phi$ over which this expression applies depends on how closely the real fields (magnetic and density fields) are described by our assumed fields.  

The precise functional form of $F(\phi)$ will also dictate the integrated flux over the feature.  If $F(\phi)$ falls significantly faster than $\phi^{-0.5}$, this could reduce the integrated flux to the point that the caustic may not be detectable. 

There is one scenario in which the fundamental description of a Faraday caustic will deviate from the one given above: the LOS magnetic field reduces to zero but does not change sign.  In this case, $B'_3(0)=0$, $B_3(z)\propto z^2$, and $\phi(z) \propto z^3$.  The Faraday spectrum will still be strongly peaked at $\phi_0$, but the spike will be symmetric in this case, with tails extending outward on both sides of $\phi_0$.

It is also worth noting how a caustic appears in $\lambda$-space.  From Eq. \eqref{eq:fourier_relation}, we see that the polarized intensity as a function of $\lambda$, $P(\lambda)$, is found by taking the Fourier transform of Eq. \eqref{eq:Fphi}.  Doing so, we find

\begin{equation}
	P(\lambda) = D e^{-2 i \lambda^2 \phi_0} \frac{(-1-i)\sqrt{\pi}}{2 \lambda}. 
\label{eq:caustic_in_lambda}
\end{equation}
Although this seems to be divergent for $\lambda \rightarrow 0$, we recall that $D$ contains $a_s$, which contains the spectral component of the Faraday spectra, $s(\lambda)$. This spectral component ensures that the total physical spectra is finite.

\section{Observing Faraday caustics}\label{sec:obs_examples}

Now that we have characterized Faraday caustics (to first order), we can discuss the expected appearance of these features in observations.  Since they occur whenever the magnetic field is aligned in the plane of the sky, we expect Faraday caustics to be a regularly occurring feature in RM synthesis observations of diffuse polarized fields.  Depending on the instrument used and the details of the experimental setup, the appearance of Faraday caustics may vary, and as we later demonstrate, may not exhibit the characteristic asymmetry described in the previous section.  Nevertheless, Faraday caustics will be present in many observations and so it is important to know what to expect in order to properly interpret Faraday spectra and learn as much as possible about the underlying magnetic fields.   

In lieu of observational data, we make use of the Hammurabi software of \citet{waelkens_simulating_2009} to provide examples of Faraday caustics in a realistic Faraday spectrum.  This software simulates the Galactic synchrotron emission based on user-defined models of the magnetic field, thermal, and cosmic-ray electron distributions.  Hammurabi was first used by \citet{sun_radio_2008} to test Galactic magnetic field models by simulating the resulting polarized foreground emission in the Galactic plane. Other examples of its use may be found in \citet{jansson_2009} who compared images of the Galactic synchrotron emission from WMAP5 with simulated emission using several existing Galactic magnetic field models from the literature, in \citet{oppermann_litmus_2010} who tested the novel LITMUS procedure that attempts to detect magnetic helicity in simulated sky maps, or in \citet{jaffe_galactic-mag-field_2010} who developed parametrized magnetic field models in the Galactic plane. We modified Hammurabi to calculate the Faraday spectrum; at each step of the integration along a given LOS, we compute the Faraday depth and the intrinsic polarized emission from within the cell.  This is stored in a three-dimensional (3D) cube with user-defined Faraday depth resolution.  

For the examples presented here, Hammurabi was run using the same simulation parameters as \citet{sun_radio_2008} for their ASS+RING magnetic field model including a stochastic component (as well as the cosmic-ray electron distribution and the thermal electron distribution of \citet{cordes_ne2001.i._2002}), except at lower angular resolution, and with higher LOS resolution (more integration steps).    

\begin{figure}
	\centering
	\subfigure{\includegraphics[width=8cm]{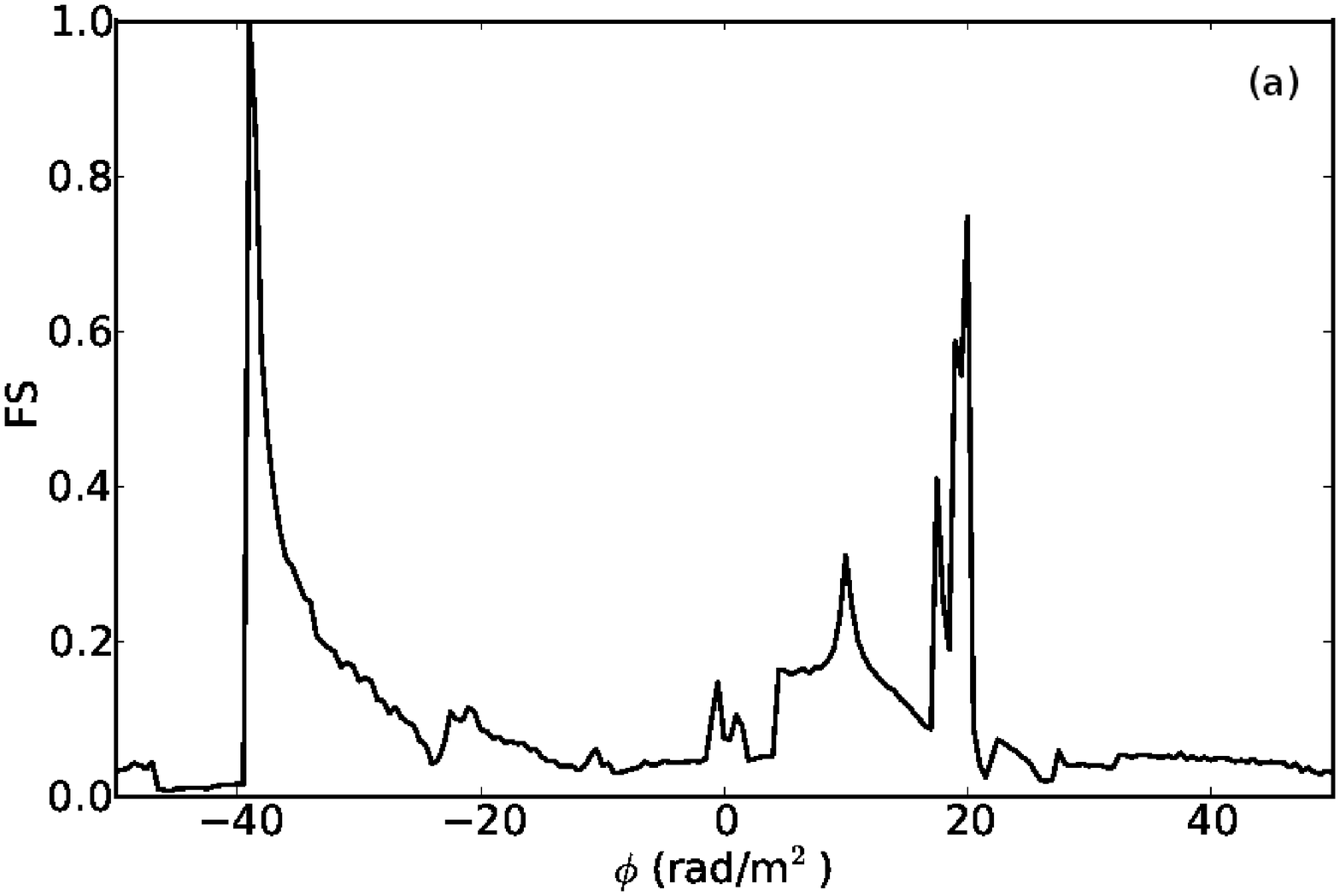}\label{fig:simul_FDF}}
	\subfigure{\includegraphics[width=8cm]{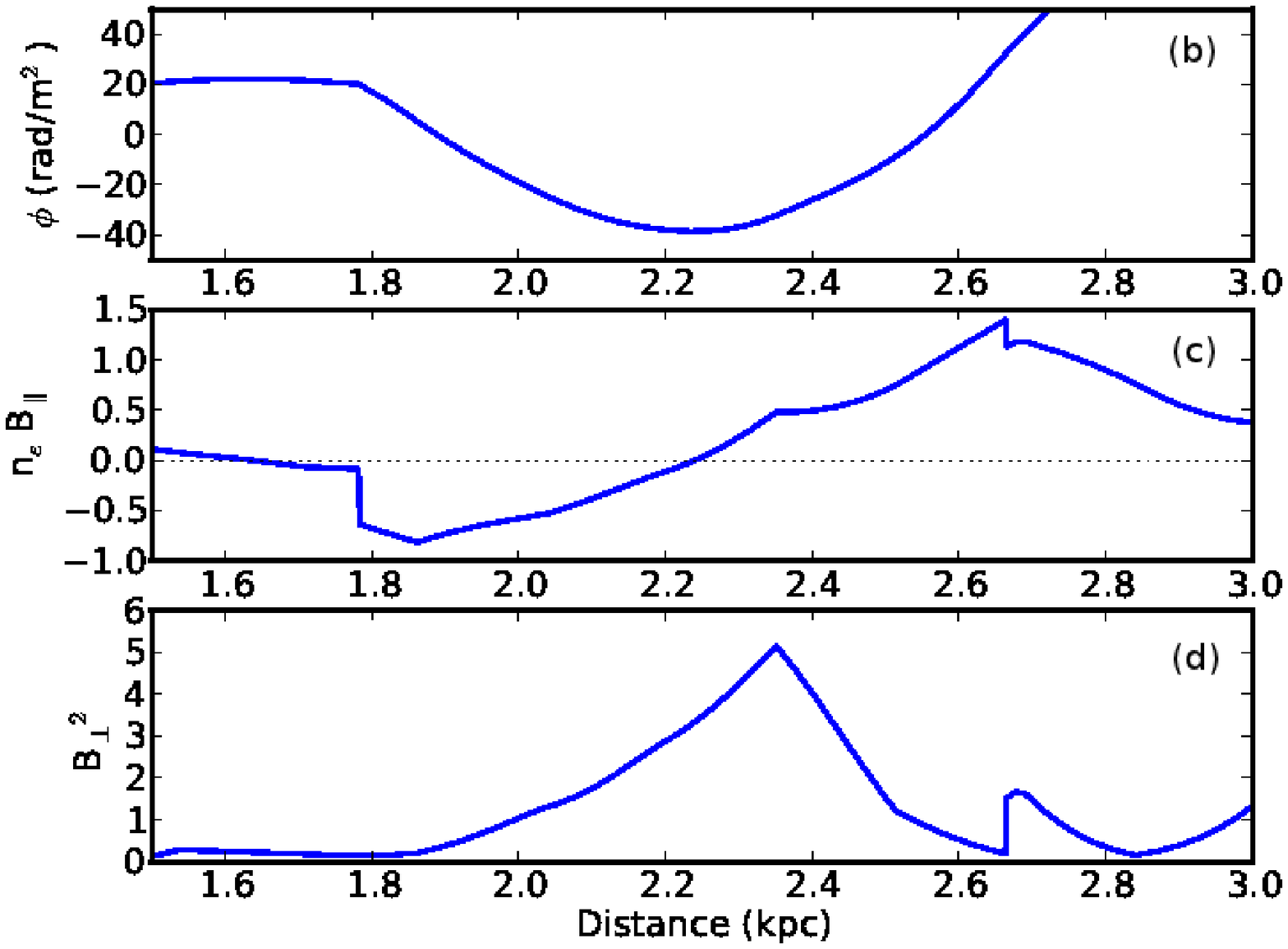}\label{fig:simul_fields}}
	\caption{A Hammurabi simulated Faraday spectrum and the associated Faraday depth and field parameters as a function of physical depth.  (a) The Faraday spectrum along the LOS at Galactic coordinates ($90^{\circ}, -15^{\circ}$) in arbitrary units. (b) Faraday depth in rad m$^{-2}$. (c) The product of the LOS magnetic field and thermal electron density, which is proportional to the Faraday depth in code internal units. (d) The perpendicular magnetic field component squared, which is proportional to the synchrotron emissivity again in code internal units.}
\end{figure}

We find that the simulated Faraday spectra typically consist of some weak diffuse emission spread over tens of rad m$^{-2}$ as well as many narrow, asymmetric ``spikes''.  In Fig. \ref{fig:simul_FDF}, we show a portion of the Faraday spectrum along the LOS at Galactic latitude and longitude ($90^{\circ}, -15^{\circ}$) that provides a clear example of the kind of features discussed in the previous section.  Along this LOS, we find a prominent caustic feature at $\phi=-39$ rad m$^{-2}$, as well as another small cluster of a few caustic features at $\phi=19$ rad m$^{-2}$.  The Faraday depth, $\vec{B}$ and $n_e$ values are shown as a function of LOS distance in Fig. \ref{fig:simul_fields}.  The product of $n_e$ and $B_{3}$ is related to the Faraday depth, and the quantity $B^2_{\perp}$ is proportional to the synchrotron emissivity.  The caustic feature at $-39$ rad m$^{-2}$ is associated with the field reversal at $z\sim2.2$ kpc, where the LOS magnetic field is zero and the synchrotron emissivity peaks.  One of the caustics in the cluster around $19$ rad m$^{-2}$ is due to the LOS magnetic field zero crossing at $z\sim 1.6$ kpc, while the others are due to magnetic field structures that are not shown.  

\begin{figure}
	\centering
	\includegraphics[width=8cm]{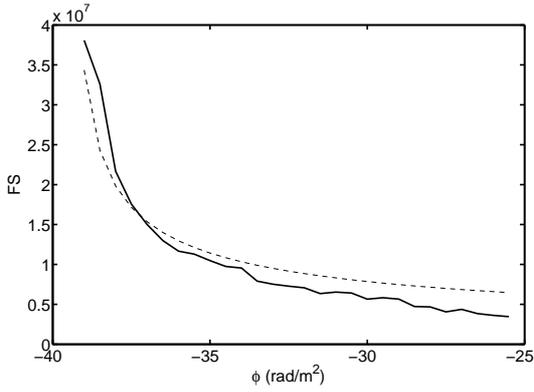}
	\caption{A portion of the simulated Faraday spectrum from Fig. \ref{fig:simul_FDF}.  This is compared to the generic form for the Faraday spectrum of a caustic, $F(\phi) \propto 1/\sqrt{\phi}$ (see Eq. \eqref{eq:Fphi}), shown as a dashed line.}
\label{fig:model_fs_fit}
\end{figure}

The major caustic shown in Figure \ref{fig:simul_FDF} at $\phi=-39$ rad m$^{-2}$ clearly exhibits the primary features described in the previous section.  It is strongly spiked, has the characteristic asymmetric shape, and is reasonably well fit by the generic $F(\phi) \propto 1 / \sqrt{\phi - \phi_0}$ form as shown in Fig. \ref{fig:model_fs_fit}.  The tail extends towards higher values of $\phi$ as a result of the positive slope of the $B_3 (z)$ distribution at $z = 2.2$ kpc.  

It should be possible to resolve particularly strong and isolated caustic features such as the one shown here, thereby allowing one to observe the characteristic asymmetry and possibly even to reconstruct the shape of the tail of the spike. The ability to resolve Faraday caustics would not only provide a means of identification, but also information about the magnetic field strength and structure. For instance, the direction of the ``tail" of the spike is dictated by the sign of $B_{3}'(z_0)$.  If $B_{3}$ changes from positive to negative, then the Faraday caustic will be extended when $\phi<\phi_0$.  In addition, if one were able to fit the tail to Eq. \eqref{eq:Fphi}, it would provide a measure of $D$ and therefore of the ratio $B_{\perp}^2(0)/\sqrt{B'_3(0)}$.

What are the necessary conditions for resolving a Faraday caustic? \citet{brentjens_faraday_2005} give the resolution, $\delta\phi$, and maximum observable scale in Faraday depth space for an RM synthesis experiment as

\begin{equation}
	\delta \phi \sim \frac{2 \sqrt{3}}{\Delta \lambda^2},
\label{eq:phi_res}
\end{equation}
\begin{equation}
	\text{max. scale} \sim \frac{\pi}{\lambda_{min}^2},
\label{eq:phi_max_scale}
\end{equation}
where $\Delta \lambda^2$ is the sampled range in $\lambda^2$ and $\lambda_{min}$ is the minimum sampled wavelength.  To reconstruct the shape of the caustic, its width must be several times $\delta \phi$.  Owing to their narrow shape, high Faraday depth resolutions will be required to resolve Faraday caustics. To achieve long ``baselines'' in $\lambda^2$-space, an experiment must extend to low frequencies.  This implies that low frequency instruments such as LOFAR will be ideally suited to observing these types of features.  

We also require that the maximum scale, given by Eqn. \ref{eq:phi_max_scale}, is larger than the resolution, $\delta\phi$, in order to have a chance of reconstructing the caustics shape.  From Eqs. \eqref{eq:phi_res} and \eqref{eq:phi_max_scale}, we find that this condition is satisfied if 

\begin{equation}   
	\frac{\nu_{max}}{\nu_{min}} > \sim 1.5.
\label{eq:freq_ratio}
\end{equation}
We consider the high-band LOFAR receivers that operate between 110 and 250 MHz.  The FWHM of the main peak of the point spread function in Faraday depth space is $\delta \phi \sim 0.5$ rad/m$^2$ and the maximum $\phi$ scale is roughly three times larger.  

\begin{figure}%
	\centering
	\subfigure{\includegraphics[width=8cm]{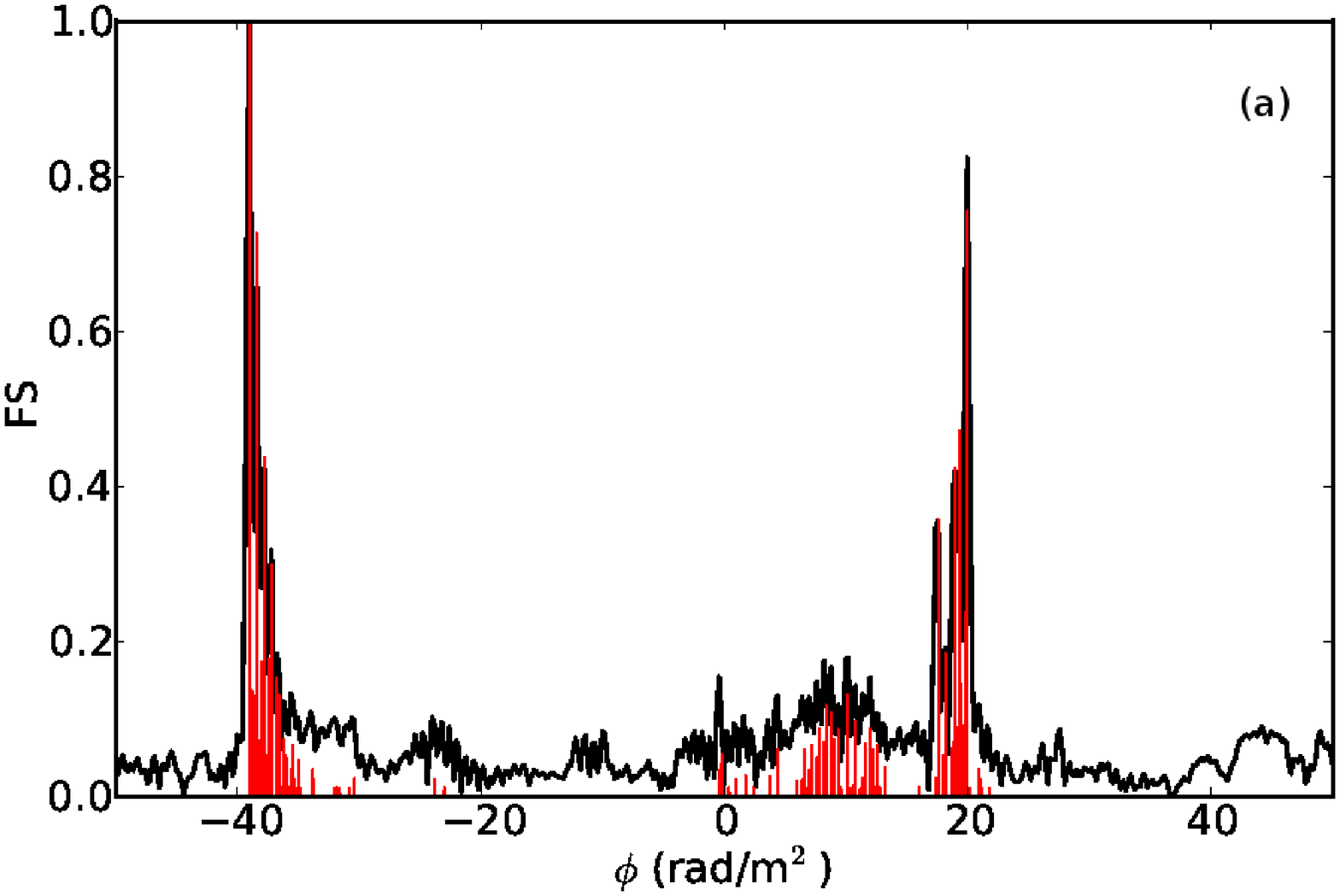}\label{fig:lofar_hba_obs}}
	\subfigure{\includegraphics[width=8cm]{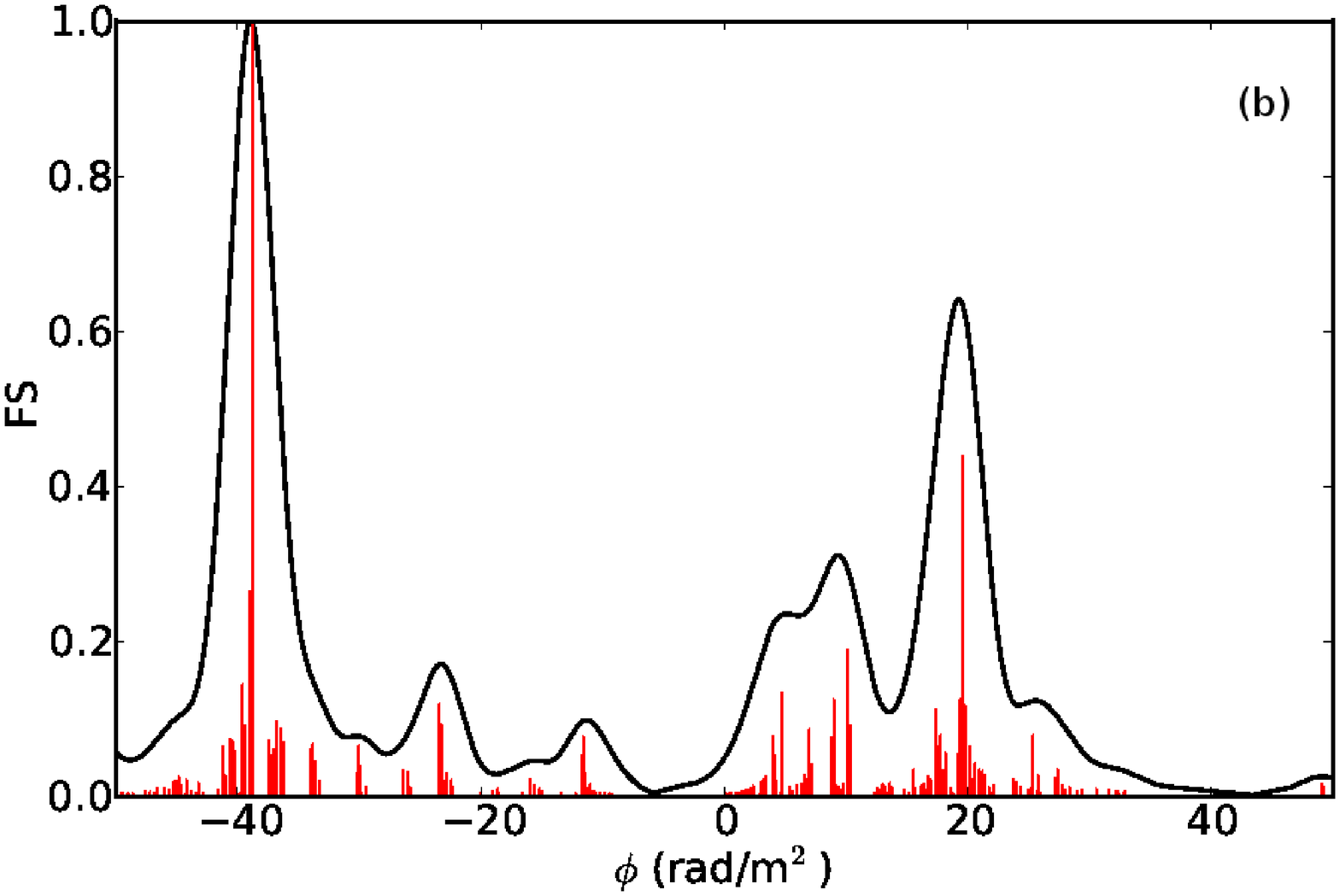}\label{fig:parkes_obs}}
	\caption{Mock observations of the Faraday spectrum shown in Fig. \ref{fig:simul_FDF}. The spectrum has been observed over two frequency ranges: (a) 110 to 250 MHz, similar to the LOFAR high band antennas and (b) 300 to 900 MHz similar to the low frequency portion of the GMIMS survey.  In each figure, the Faraday spectrum has been deconvolved using RMCLEAN.  The black line represents the restored spectrum, while the RMCLEAN components are shown as vertical grey lines (red in the online version).} 
	\label{fig:two_mock_obs}%
\end{figure}

As an example of what we can expect caustics to look like in real data, we can ``observe'' our simulated Faraday spectra by evaluating its Fourier transform in $\lambda^2$-space, sampling at some set of frequencies, and then Fourier transforming back to $\phi$-space.  Figures \ref{fig:lofar_hba_obs} and \ref{fig:parkes_obs} show the Faraday spectrum of Fig. \ref{fig:simul_FDF} as ``observed'' using two different frequency ranges.  The frequency range used for Fig. \ref{fig:lofar_hba_obs} is 110-250 MHz, which is similar to that for the LOFAR high band antennas.  The frequency range used for Fig. \ref{fig:parkes_obs} is from 300-900 MHz, which is similar to the low frequency portion of the GMIMS survey \citep{wolleben_gmims_2009}.  We include 512 frequency samples over each range.  Most next generation instruments will be able to observe using many more frequency channels than this.  In each figure, we show the Faraday spectrum after deconvolution using our own RMCLEAN software that implements a procedure similar to that described by \citet{heald_westerbork_2009}. The RMCLEAN components are shown in red.

In the ``LOFAR'' reconstructed spectrum, the stronger caustic is reasonably well recovered.  The shape of the tail is not exactly as it is in the original spectrum, but the asymmetry is clearly visible.  The small cluster of caustics at $\phi \sim 20$ rad/m$^2$ is resolved showing the three distinct peaks.  The same cannot be said of the ``GMIMS'' spectrum.  Owing to its lower resolution ($\delta \phi \sim 5$ rad/m$^2$ compared to $\sim 0.5$ rad/m$^2$ for the ``LOFAR'' observation), the individual peaks at $\phi \sim 20$ rad/m$^2$ are not resolved.  The shape of the large caustic is completely lost, even though the tail of the feature extends over $15$ rad/m$^2$ in the original spectrum.  No asymmetry is present even in the distribution of CLEAN components.  

Observations with lower $\phi$-space resolution, i.e. a smaller $\delta \lambda^2$, will of course fair worse than the above examples.  We consider the Australian Square Kilometre Array Pathfinder (ASKAP) POSSUM survey \citep{gaensler_possum_2010} that will operate over the range of frequencies from 700 MHz to 1.8 GHz.  The $\phi$-space resolution will be approximately 22 rad/m$^2$, which is lower than the examples included here.  While caustics will likely still be detected by the POSSUM survey, and other experiments operating at higher frequencies, they will simply appear as unstructured point sources.   
  
Having described the appearance of individual caustic features in an observed Faraday spectrum, we now investigate how these are distributed throughout the 3D Faraday spectral cube.  Faraday caustics are generated at the boundaries of regions with opposing LOS magnetic field polarity. This boundary forms a surface in physical space. If the magnetic field distribution is well ordered, this surface will be smooth and map onto a connected surface in $\phi$-space. Therefore, Faraday caustic surfaces will reflect boundaries between regions of opposite LOS magnetic field polarity.  

As mentioned in the previous section, there are a symmetric variety of Faraday caustics that appear as a result of the LOS magnetic field approaching zero tangentially without changing sign.  These features will appear when the boundary between regions of opposite LOS magnetic field polarity is aligned along the LOS, i.e. where a Faraday caustic surface folds over.  

In Figs. \ref{fig:movie_slices1} and \ref{fig:movie_slices2}, shown in Appendix \ref{sec:cube_slices}, we show six slices from a Hammurabi simulated Faraday spectrum at different Galactic latitudes. The top panel of each slice shows the Faraday spectrum as a function of Galactic longitude and Faraday depth, while the bottom panel shows the LOS magnetic field strength as a function of Galactic longitude and LOS distance.  The scale of the magnetic field strength has been compressed to highlight the regions of opposite polarity.  White regions indicate a negative LOS magnetic field, black regions a positive field, and the colored sections highlight the boundaries where $B_3 \approx 0$.  

In each slice, we can see that the smoothly connected boundaries between regions of opposite LOS magnetic field polarity produce connected lines of Faraday caustics in $\phi$-space.  The caustics can be seen clearly in the Faraday spectrum, appearing as sharp, bright lines with faint tails extending to one side.  Watching how the lines move from frame to frame, we can get a sense of how the caustics are distributed in 3D, appearing as a surface of bright polarized emission.  In Figure \ref{fig:movie_slices1}, at Galactic latitude $-117.6^{\circ}$,  a continuous line of caustic features extends from Galactic longitude $15^\circ$ and $\phi \approx -5$ rad m$^{-2}$ to Galactic longitude $60^\circ$ and $\phi \approx 0$ rad m$^{-2}$.  This surface reflects the $B_{3}=0$ boundary at $z\approx 0.2$ kpc.  This continuous line remains dominant until frame c, where the polarized emission has shifted to $\phi=0$ rad m$^{-2}$.  At this point, a second $B_{3}=0$ surface, where the LOS magnetic field changes from negative to positive, is forming between $4$ and $6$ kpc.  The $B_{3}=0$ surface between longitudes -15$^\circ$ and 15$^\circ$ and at $z\sim 0.6$ kpc produces the line of Faraday caustics between $\phi \sim 0$ and $5$ rad m$^{-2}$.   

In Figure \ref{fig:movie_slices2}, the rightmost $B_{3}=0$ surface produces the dominant lines of Faraday caustics extending to positive $\phi$ values in the Faraday spectra.  The Faraday caustic surface associated with the leftmost $B_{3}=0$ surface is now at $\phi=0$ rad/m$^2$.  We note that the tails of the caustic spikes extend between the two surfaces since the LOS magnetic field changes from negative to positive causing one set of caustics, and from positive to negative for the other.

\begin{figure}
	\centering
	\includegraphics[width=10cm]{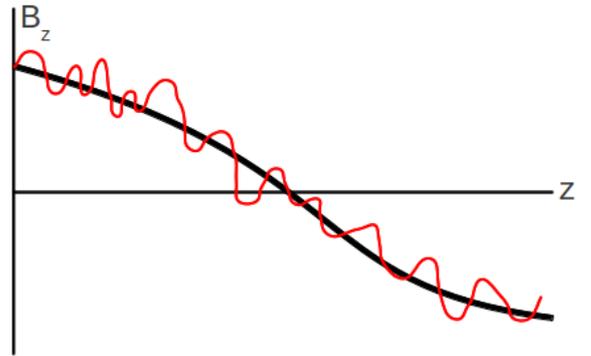}
	\caption{A diagram illustrating how a small-scale fluctuation about a larger scale magnetic field variation may split a single Faraday caustic (occurring at the zero crossings) into several smaller caustics.  An observation will not be sensitive to fluctuations below a given scale length that depends on the resolution of the experiment.}
	\label{fig:uni_and_rand_bfield}
\end{figure} 

Depending on the smoothness of the magnetic field variation, and the resolution of a given experiment, the caustic surface may consist of a single, well-defined peak or a tight cluster of smaller spikes.  We draw attention to the LOS magnetic field sketched in Fig. \ref{fig:uni_and_rand_bfield}, where the black line represents a large-scale variation in the magnetic field, while the grey curve (red in the online version) represents small-scale fluctuations about this distribution.  For a smooth variation such as that represented by the black line, a single Faraday caustic will be present.  In the case of the red curve, three caustics will be created close to one another in $\phi$-space. The limited resolution of a given experiment acts to effectively smooth the magnetic field distribution by setting a minimum length scale below which the experiment is insensitive to variations.  In other words, the $\delta \phi$ resolution also ensures a finite resolution in $z$-space.  Hence, we expect Faraday caustic surfaces to be either sharp, smoothly connected features in the Faraday spectra, or a more fractured boundary depending on the scale of fluctuation of the magnetic field and the resolution, $\delta \phi$.

\section{Distribution of Faraday caustics in a Gaussian random magnetic field}\label{sec:caustics_in_gauss_field}

We have just shown how Faraday caustics can provide information about the structure of the LOS magnetic field from their shape and 3D distribution.  It is variation and turbulence in the magnetic field that leads to caustics; after all, there would be no such features in a purely uniform magnetic field.  It may be possible to recover information about the statistical properties of the underlying magnetic field in a source by studying the distribution of caustics within its Faraday spectrum.  We therefore now investigate the production of Faraday caustics by a Gaussian random magnetic field and calculate their probability distribution as a function of strength, i.e. the luminosity function. For detailed derivations of all results in this section, we refer to Appendix \ref{sec:probability_derivations}.

We assume that the magnetic field distribution is a Gaussian random field with given covariance matrix $M$.  For a Gaussian random field, the covariance matrix is also equal to the correlation tensor. The translationally invariant magnetic correlation tensor for homogeneous and isotropic magnetic turbulence is 
\begin{align}
&M_{ij}(r)=\left\langle B_i(\textbf{x}) B_j(\textbf{x} + \textbf{r}) \right\rangle \notag\\
&M_{ij}(r)=M_{N}(r)\delta_{ij}+\big(M_{L}(r)-M_{N}(r)\big)\frac{r_{i}r_{j}}{r^2} \notag \\ 
&\phantom{M_{ij}(r)=} +M_{H}(r)\epsilon_{ijm}r_{m} \ \text{in real space} \notag\\
&\hat{M}_{ij}(k)=\hat{M}_{N}(k)(\delta_{ij}-\frac{k_{i}k_{j}}{k^{2}})\notag\\
&\phantom{\hat{M}_{ij}(k)=}-i\epsilon_{ijm} \hat{H}(k) \frac{k_{m}}{k} \ \text{in Fourier space}  
\label{eq:magnetic_correlation_tensor}
\end{align}
\citep[e.g.]{1999subramanian} where the longitudinal, normal, and helical auto-correlation functions, $M_L (r)$, $M_N (r)$, and $M_H (r)$, respectively, only depend on the magnitude of the separation between two points in space.  We note that $M_N (k)$ can be expressed in terms of the one-dimensional (1D) magnetic field power spectrum $\epsilon_B(k)$ as

\begin{equation}
	\epsilon_B(k) = \frac{k^2 M_N(k)}{8 \pi}. 
	\label{eq:eb_to_MN}
\end{equation}

The probability of measuring an integrated intensity of $\mathcal{F}$ at some location $z_0$, given that there is a Faraday caustic at $z_0$ and given the magnetic covariance matrix $M$, is
\begin{equation}
	P(\mathcal{F}|\text{caustic},M) = \frac{P(\mathcal{F},\text{caustic}|M)}{P(\text{caustic}|M)}.
\label{eq:prob_f_setup}
\end{equation}   
The denominator, $P(\text{caustic}|M)$, is fairly straightforward to compute
\begin{equation}
	P(\text{caustic}|M) = \int \; \mathcal{D}B \;\delta(B_{3}(z)) \;\mathcal{G}(B,M),
\label{eq:prob_spike_setup}
\end{equation}
where $\mathcal{D}B$ indicates a phase space integral over all possible B-field configurations.  In addition, $\mathcal{G}(\vec{B},M)$, a Gaussian probability density function over this phase space, is 

\begin{equation}
	\mathcal{G}(B,M) = \frac{1}{\sqrt{\left|2 \pi M\right|}} e^{-\frac{1}{2}B^{\dagger}M^{-1}B},
\label{eq:gaussian_def}
\end{equation}
where vertical bars indicate that the determinant is to be taken. Here, the notation $A^{\dagger}B$ implies a multidimensional scalar product such that 
\begin{equation}
	A^{\dagger}B = \sum_i \int dx^3 \; A^*_i(x)B_i(x).
\label{eq:}
\end{equation}
The sum is over the field components and the integral is over physical space.

We find that the probability of a caustic occurring (i.e. that $B_3(z_0)=0$) is   

\begin{equation}
	P(\text{caustic}|M) = \sqrt{\frac{2 \pi}{M_{33}(0)}},
\label{eq:prob_spike}
\end{equation}
where $M_{33}$ is the component of $M$ that describes the two-point correlation between $z$ components of the magnetic field.

The calculation of the numerator is significantly more involved and as we describe in Appendix \ref{sec:probability_derivations}, Eq. \eqref{eq:prob_f_setup} can only be computed analytically in the limit where $\mathcal{F}$ is large (we define this more clearly below).  In this limit, we obtain the result

\begin{equation}
	P(\mathcal{F} | \text{caustic}, M) = \frac{448 a_s^2 \delta \phi^3}{a_1^2}\frac{M_N^2(0)}{\sqrt{2 \pi \left|M''_{33}(0)\right|}} \frac{1}{\mathcal{F}^3}.
\label{eq:prob_f_result}
\end{equation}
We find that, over a limited range in $\mathcal{F}$, the distribution of the integrated flux of caustics will be proportional to $\mathcal{F}^{-3}$.  We note that this indicates that the number of caustics per logarithmic interval of $\mathcal{F}$ is $\propto \mathcal{F}^{-2}$ and thus that the Faraday caustic flux per logarithmic interval of $\mathcal{F}$ goes as $\mathcal{F}^{-1}$, which is finite as $\mathcal{F} \rightarrow \infty$.  

The fore-factors in Eq. \eqref{eq:prob_f_result} depend on the properties of the magnetic field and the turbulent flow of the fluid in which they are situated. The term $M_N(0)$ is simply the total magnetic energy density.  We also introduce the \emph{Taylor microscale}, $l_T$, which is defined by

\begin{equation}
	\frac{d^2 C(z)}{dz^2}\bigg|_{z=0} = \frac{2}{l_T^2}
	\label{eq:talor_microscale}
\end{equation}
\citep[see e.g.][sec. 6.4]{tennekes_lumley_turbulence_1972}, where $C(z)$ is the covariance in $z$ normalized to unity at $z=0$.  This length scale represents the largest scale on which dissipation is important in a turbulent flow.  The second derivative of the LOS component of the magnetic correlation tensor, $M''_{33}(0)$, is proportional to $\epsilon_B/l_T^2$.  We can therefore rewrite Eq. \eqref{eq:prob_f_result} as

\begin{equation}
	P(\mathcal{F} | \text{caustic}, M) = \frac{448 a_s^2 \delta \phi^3}{a_1^2}\frac{\epsilon_B^{3/2} l_T}{\sqrt{4 \pi}} \frac{1}{\mathcal{F}^3},
\label{eq:prob_f_w_taylor_scale}
\end{equation}
where $\epsilon_B$ is the total magnetic energy density.  

By measuring the luminosity function of Faraday caustics in the $\mathcal{F}^{-3}$ regime, we can see that one is in principle able to measure the Taylor microscale in a turbulent medium.  \citet{fletcher_canals_2007} show that this length scale is also measurable by observing the mean separation between so-called ``depolarization canals,'' which are lines of zero polarized intensity in a diffuse polarized field.  These canals have been observed in maps of diffuse polarized Galactic emission by, e.g. \citet{uyaniker_medlatsurvey_1998}, \citet{haverkorn_ism-structure_2000}, \citet{gaensler_inner-galaxy-pol_2001}, and \citet{reich_emls_2004}. Observations of depolarization canals and Faraday caustics in tandem may be a powerful tool for studying the turbulent properties in, e.g. the ISM.

A rather important caveat, applicable to both Faraday caustics and depolarization canals, is that without sufficiently high resolution one will not be able to measure this scale.  In the case of caustics, as discussed in the previous section, the effect of small-scale fluctuations is to split a single Faraday caustic into a tight bundle of spikes.  Without enough resolution, an observer may simply count such a bundle as one large spike.  This has the effect of altering the normalization of the number distribution of caustics in $\mathcal{F}$, which would lead to an incorrect measurement of $l_T$.  In principle, the same effect applies to depolarization canals, where what appears to be a single canal of net zero polarized intensity may be a fine network of such canals.  As described previously, the effect of finite resolution is to smooth the quantity being measured, e.g. the LOS magnetic field distribution in the case of Faraday caustics, thereby throwing away information on any smaller scales.  If the Taylor microscale is smaller than the smoothing scale set by our experimental resolution, then it will not be measurable. We suspect that this may be the reason why the Taylor scale of the Milky Way is measured to be much larger than expected in the example presented by \citet{fletcher_canals_2007}.

The result given in Eq. \eqref{eq:prob_f_w_taylor_scale} is valid for strong spikes, i.e. when $B'_3(0)$ is small, but we expect that below some value of $\mathcal{F}=\mathcal{F}_{low}$ the distribution will flatten.  This is required to ensure that the integral of $P(\mathcal{F}|\text{caustic},M)$ is finite for $\mathcal{F} \rightarrow 0$, but we can also see that this will be the case by considering the approximation applied during our derivation.

To permit us to compute the probability analytically, we neglect an exponential term in Eq. \eqref{eq:now_we_need_perturbation_theory}.  This term is $\exp(-\mu K B^4_{\perp}/\mathcal{F})$, where $K=16 a_s^2 \delta \phi / a_1$. This is, of course, only valid if the argument of the exponent is much less than one.  The approximation breaks down when this term approaches unity, which leads to the condition for $\mathcal{F}_{low}$

\begin{align}
	\mathcal{F}_{low} & \approx \left\langle B_{\perp}^2\right\rangle \sqrt{\frac{K}{\left\langle B'_3 \right\rangle}} \\
	& \approx K^{\frac{1}{2}} \epsilon_B^{\frac{3}{4}} \lambda_B^{-\frac{1}{2}}, 
\label{eq:low_lim_condition}
\end{align}
where 
\begin{equation}
	\epsilon_B = \int_0^{\infty} dk \epsilon_B(k)
\label{eq:mag_energy_density}
\end{equation}
is the average magnetic energy density, $\epsilon_B(k)$ is the 1D magnetic power spectrum, and
\begin{equation}
	\lambda_B = \frac{\pi \int_0^{\infty} dk \epsilon_B(k)/k}{\int_0^{\infty} dk \epsilon_B(k)}
\label{eq:mag_correlation_length}
\end{equation}
is the magnetic correlation length.

If we wish to evaluate Eq. \eqref{eq:now_we_need_perturbation_theory} in all regimes, we can no longer neglect this term. While this full calculation is beyond the scope of this paper, we can consider what effects the inclusion of this term might have on the result. To proceed without approximations, we would use a pertubative expansion in Feynman diagrams as is used in quantum field theory.  The first term in such an expansion would be negative and $\propto \mathcal{F}^{-2}$, the second positive and $\propto \mathcal{F}^{-1}$, and so on.  The summation of these terms results in a turnover of the spectrum near $\mathcal{F}_{low}$.

We also expect the distribution to steepen at larger values of $\mathcal{F}$ because of depolarization.  As discussed above, large values of $\mathcal{F}$ imply that the LOS magnetic field changes over long distances.  As this happens, it is increasingly more likely that the position angle of the polarized emission at the values of $z$ that contribute to the polarized intensity at a single value of $\phi$ will be uncorrelated resulting in a reduction in the integrated intensity because of depolarization.  This reduces the number of caustic features at high values of $\mathcal{F}$. 

To investigate the conditions for a steepening of the probability distribution of caustics, we consider the correlation of the sky plane magnetic field as a function of LOS distance

\begin{equation}
	\mathcal{R} = \frac{\left\langle B_{\perp}(\overline{z}) B_{\perp}(-\overline{z})\right\rangle}{\left\langle B^2_{\perp}(0)\right\rangle},
\label{eq:bfield_correlation_ratio_setup}
\end{equation}  
where the average is over all possible field configurations.  The procedure for computing these expectation values is similar to that used in Appendix \ref{sec:probability_derivations} to compute $P(\mathcal{F} | caustic, M)$ and therefore not included here.  The result is
\begin{equation}
	\mathcal{R} = \frac{3 M_N(2 \overline{z})}{10 M_N(0)} + \frac{7 M^2_N(\overline{z})}{10 M^2_N(0)}.
\label{eq:bfield_correlation_ratio_result}
\end{equation}
A small value of $\mathcal{R}$ indicates that, on average, the sky plane magnetic fields are uncorrelated.  When this is true, the luminosity function is steeper than $\mathcal{F}^{-3}$.

As an example of estimating $\mathcal{F}_{low}$ and $\mathcal{R}$, we model the magnetic power spectrum as a broken power law

\begin{equation}
	\epsilon_B(k) = \epsilon_0 \left(\frac{k}{k_0}\right)^{b} \left[1 + \left(\frac{k}{k_0}\right)^2\right]^{-(\frac{a+b}{2})}.
\label{eq:broken_power_law}
\end{equation}
For such a power spectrum, the luminosity function turns over at
\begin{equation}
	\mathcal{F}_{low} = \pi^2 \sqrt{K} \left( \frac{\epsilon_0 k_0^{5/3}}{2} \right)^{\frac{3}{4}} \mathcal{B}\left(\frac{a-1}{2}, \frac{b+1}{2}\right)^{\frac{5}{4}} \mathcal{B}\left(\frac{a}{2}, \frac{b}{2}\right)^{-\frac{1}{2}},
\label{eq:f_low_for_bpl}
\end{equation}
where $\mathcal{B}(x)$ is the Beta function.

\begin{figure}%
	\centering
	\includegraphics[width=8cm]{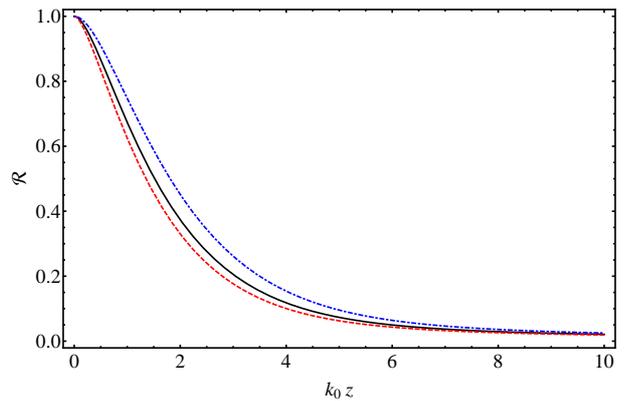}
	\caption{The normalized correlation of the sky plane magnetic field as a function of $k_0 z$ for a few values of the power law index $a$.  Dashed red $a=4/3$, solid black $a=5/3$, dot-dashed blue $a=7/3$.} 
	\label{fig:R_v_kz}%
\end{figure}

We can also compute $\mathcal{R}$ because
\begin{equation}
	M_N(z) = \int dk \hat{M}_N(k) e^{-i k z}
\label{eq:MN_fourier}
\end{equation}
and 
\begin{equation}
	\hat{M}_N(k) = \frac{8 \pi^3 \epsilon_B(k)}{k^2}.  
\label{eq:MNk_to_epsBk}
\end{equation}
The resulting form of $\mathcal{R}(k_0 z)$ is shown for a few representative values of $a$ in Fig. \ref{fig:R_v_kz}.  For a Kolmogorov spectrum, $a=5/3$. In each case, $b=2$.  We note that $\mathcal{R}$ is independent of the parameter $\epsilon_0$.

\begin{figure}%
	\centering
	\includegraphics[width=8cm]{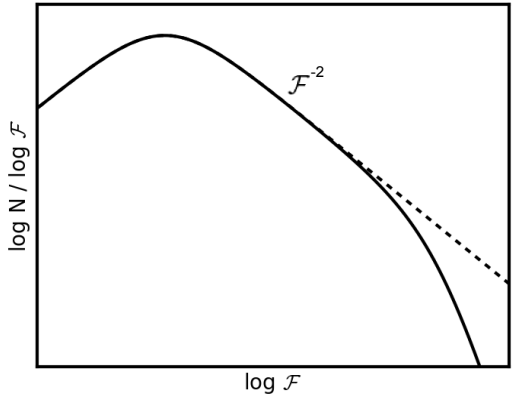}
	\caption{A sketch of the number distribution of caustic spikes per logarithmic interval of $\mathcal{F}$ showing the limits to the $\mathcal{F}^{-2}$ regime and the rough shape of the distribution beyond these limits.  At high values of $\mathcal{F}$, the distribution steepens owing to depolarization effects, while at lower $\mathcal{F}$ values the distribution turns over because the total number of caustics is finite.  We refer to the text for more details.} 
	\label{fig:prob_dist_sketch}%
\end{figure}

A calculation of the precise shape of the distribution for low and high values of $\mathcal{F}$ is beyond the scope of this paper.  A sketch summarizing the expected shape of the probability distribution, represented as the number of caustics per logarithmic scale of $\mathcal{F}$ as a function of log $\mathcal{F}$, is shown in Fig. \ref{fig:prob_dist_sketch}. The $\mathcal{F}^{-2}$ regime is valid given the assumptions outlined above for sufficiently strong caustics and until depolarization becomes important, as described above.  We expect that the exact shape of the distribution at and below $\mathcal{F}_{low}$ will be more strongly dependent on the magnetic field statistics because weak caustic features depend more strongly on $\left|B_{\perp}(0)\right|$.

In summary, we have calculated the probability density function of caustics with a particular strength in a Gaussian random magnetic field.  This result illustrates the type of analysis that is possible and represents the simplest reference case for future observations of the luminosity function.  We caution that the result has a limited scope for several reasons. For one, we have assumed an unrealistic magnetic field distribution.  This was done for simplicity, but also because if one is unaware of any statistical properties of the magnetic field other than the two point correlation, the assumption of a Gaussian field is the most appropriate starting point.  Any further knowledge would always be included by expanding around the Gaussian case. While we expect our result in the $\mathcal{F}^{-3}$ regime to be only weakly dependent on the magnetic field statistics, realistic magnetic field distributions are likely to be significantly non-Gaussian.  We have not investigated the degree to which this will effect our results. Secondly, the result in Eq. \eqref{eq:prob_f_w_taylor_scale} is only valid over a limited range of $\mathcal{F}$ values.  The precise extent of this range is unclear at the moment, although we have presented some estimates.  It is possible that the flattening and steepening regimes that we describe are near one another or even overlapping.  A more detailed investigation into the flattening regime, for example, would require much more complicated calculations and, because this regime depends strongly on the magnetic field statistics, a more realistic model.  

Lastly, Eq. \eqref{eq:prob_f_w_taylor_scale} is truly only valid for very high resolution observations or simulated data.  As discussed above, the result will change depending on $\phi$-space resolution.  In our treatment, we simply count the number of zero-point crossings of the LOS magnetic field.  In observations, two or more of these events in physical space may occur at nearly the same $\phi$ location and may therefore be counted as a single caustic.  This would have the effect of changing the overall normalization of the distribution (the total number of caustics would be reduced), but the $\mathcal{F}^{-3}$ prediction should still be valid. Some additional analysis of simulations would be helpful in assessing the extent of this effect.

\section{Discussion and conclusions}\label{sec:conclusions}

We have introduced sharp, asymmetric features in the Faraday spectrum of a diffuse polarized source that appear as a result of a change in the direction of the LOS magnetic field.  We have called these features Faraday caustics, and derived their fundamental properties by considering a simple magnetic field distribution near the point where $B_3=0$.  The general properties of a Faraday caustic are:

\begin{itemize}
	\item They are singularities in the Faraday spectrum. The polarized intensity peaks sharply at the Faraday depth $\phi_0$ that maps to the physical depth $z_0$, where $B_3(z_0)=0$.
	\item The Faraday spectrum extends only to one side of the singularity.  The direction in which the tail extends depends on the slope of the $B_3(z)$ distribution.  If $B_3'(z) > 0$, then the tail extends to $\phi > \phi_0$.  An exception occurs when $B_3(z)$ approaches zero tangentially without changing sign.  In this case, the caustic is still strongly peaked, but symmetric.
	\item Strong caustics will predominantly be created when the LOS magnetic field changes over large physical distances i.e. when $B_3'(z_0)$ is small.  In this case, the polarized emission from even a relatively weakly emitting region (in physical space) can appear as bright point sources in the Faraday spectrum because the emission over a long LOS distance can pile up over a narrow range of Faraday depths.
	\item In a 3D Faraday spectrum, Faraday caustics will appear as continuous sheets or surfaces.  These surfaces are related to boundaries between regions of opposite LOS magnetic field polarity.  
	\item The effect of small-scale fluctuations about larger-scale variations will be to split a single large Faraday caustic into a series of smaller, closely packed caustics.  Without sufficiently high resolution, this splitting will not be observed.  In this way, the finite resolution of an experiment acts to smooth the $B_3(z)$ distribution.  
\end{itemize}     

We find that important information about the magnetic field distribution can be recovered by observing these features.  Reversals of the LOS magnetic field can be identified and if the asymmetry of the Faraday caustics is recovered, one will even be able to indicate in which direction the reversal occurs.  We have also shown that there may be great value in studying the statistical properties of Faraday caustics.  We have computed the probability distribution of the polarized intensity of Faraday caustics in a Gaussian random magnetic field.  The number of caustics per unit $\mathcal{F}$, which is the integrated intensity of a single caustic, goes as $\mathcal{F}^{-3}$.  There is a lower $\mathcal{F}$ limit at which the distribution flattens and turns over, and a high $\mathcal{F}$ limit where the distribution steepens because of depolarization effects.  More work is needed to more clearly define these limits, but this should be a fruitful endeavor because we expect that these limits will depend on the statistical properties of the magnetic field distribution.  We show that if one has sufficiently high resolution and measures this distribution in the $\mathcal{F}^{-3}$ regime one can recover the Taylor microscale.  This scale is the largest distance over which dissipation is important in a turbulent flow.      

Faraday caustics will not be the only source of point-like structure in the Faraday spectrum of a diffuse source. Point sources in Faraday space will also occur as a result of a discrete source along the LOS, such as a pulsar or any strong, compact emitter.  These discrete sources will also appear as compact sources in the sky plane and will often have associated structure in total intensity making them easy to distinguish from Faraday caustics.  If a caustic is resolved, the asymmetric shape will of course be an obvious means of identification.  In the case of structures more extended in Faraday depth space than the largest measured scale (c.f. Eq. \eqref{eq:phi_max_scale}), the extended nature of the source will not be recovered and the edges of the structure will appear to be point-like (see e.g. Fig. 2 in \citet{frick_wavelet-based_2010}).

It should be noted that caustics may already have been observed.  Though we have not analyzed the data directly, the published data from the Perseus field \citep{de_bruyn_diffuse_2005, brentjens_wide_2010} contains many point-like features, some of which seem to be connected in what may be the Faraday caustic sheets we discuss above.  A re-analysis of these data may provide the first observational evidence of Faraday caustics.

New observational techniques and imaging algorithms may be needed in order to optimally observe Faraday caustics.  When investigating the effectiveness of a wavelet-based approach to RM synthesis, \citet{frick_rmsynthesis_galaxies_2011} noted sharp, point-like features in a Faraday spectrum produced by the small-scale fluctuations included in their magnetic field distribution model.  They also first showed the characteristic asymmetric caustic profile resulting from a large-scale magnetic field reversal in their Fig. 5.  Neither of these caustic features were discussed in detail, or identified as caustics, but the authors showed that wavelet-based RM synthesis is effective at separating small and large scale features in the Faraday spectrum and as a result the wavelet-based algorithm outlined in \citet{frick_wavelet-based_2010} may be a useful tool for observing Faraday caustics.  In addition, matched filter algorithms could be helpful if one wants to more accurately reconstruct the shape of the ``tails'', which contain most of the total polarized flux and information about the magnetic field distribution.


\begin{acknowledgements}
This research was performed in the framework of the DFG Forschergruppe 1254 ”Magnetisation of Interstellar and Intergalactic Media: The Prospects of Low-Frequency Radio Observations”. We acknowledge Rainer Beck, Michiel Brentjens, Andrew Fletcher, Ue-Li Pen, Wolfgang Reich, and Xiaohui Sun for their helpful ideas and insights.  We thank Marco Selig, Helin Weingartner, and Maximilian Uhlig for many useful discussions, and especially Niels Oppermann for his careful reading of our manuscripts.  We also gratefully acknowledge the insightful remarks of the referee, Steve Spangler. Some of the results in this paper have been derived using the HEALPix package \citep{gorski_healpix_2005}.
\end{acknowledgements}

\bibliographystyle{aa}
\bibliography{rmsynth_refs}

\appendix
\onecolumn
\section{Derivation of the probability distribution for caustics}\label{sec:probability_derivations}

We derive the probability of measuring a spike with an integrated intensity $\mathcal{F}$ given that the spike is due to a caustic (i.e. $B_3 = 0$) and a magnetic covariance matrix $M$.  We limit our discussion to the regime where Eq. \eqref{eq:int_f_lowu} adequately describes the integrated flux of the caustic.

In Sect. \ref{sec:basic_features}, we found that the integrated polarized intensity of a caustic is $\mathcal{F}=4 a_{s} \sqrt{\delta \phi} \left|B_{\perp}(0)\right|^2/\sqrt{B_{3}' (0)a_{1}}$. We assume a Gaussian magnetic field distribution, $\mathcal{G}(B,M)$. The probability distribution of integrated intensities $\mathcal{F}$, given that $B_3=0$ and the magnetic correlation tensor is $M$ is
\begin{equation}
	P(\mathcal{F}|\text{caustic},M) = \frac{P(\mathcal{F},\text{caustic}|M)}{P(\text{caustic}|M)}. 
\label{eq:prob_start}
\end{equation}
The denominator, which is the probability that a caustic occurs in a Gaussian random magnetic field, is relatively straightforward to compute. We impose the condition that $B_{3}(0) = 0$ using a delta function in order to compute the probability

\begin{equation}
	P(\text{caustic}|M) = \int \; \mathcal{D}B \;\delta(B_{3}(0)) \; \mathcal{G}(B,M),
\label{eq:denom_step1}
\end{equation}
where the integral is a path-integral over all possible realizations of the magnetic field.  We then replace the delta function using the Fourier transform
\begin{equation}
	P(\text{caustic}|M) = \frac{1}{\sqrt{\left|2 \pi M\right|}}\int \: \mathcal{D}B \int d\eta  \: e^{2\pi i \eta B_3} \: e^{-\frac{1}{2} B^{\dagger} M^{-1} B}.
\label{eq:denom_step2}
\end{equation}
We now replace $\eta B_3$ with $\eta \delta_{i3} \vec{B}$ and after completing the square we have

\begin{align}
	P(\text{caustic}|M) = &\frac{1}{\left|2 \pi M\right|} \int \: \mathcal{D}B \int d\eta  \: \exp \bigg[-\frac{1}{2} \left(B - J^\dagger M\right)^{\dagger} M^{-1} \left(B - J^\dagger M\right) + \frac{1}{2}J^\dagger M J\bigg],
\label{eq:denom_step3}
\end{align}
where $J^\dagger=2 \pi i \eta \delta_{iz} \delta(z)$.  The $\vec{B}$ integral is the integral of a Gaussian function, which gives a factor of $\sqrt{\left|2 \pi M \right|}$.  The $\eta$ integral is also a Gaussian integral.  After integration, the result is

\begin{equation}
	P(\text{caustic}|M) = \sqrt{\frac{2 \pi}{M_{33}(0)}}. 
\label{eq:prob_spike_appendix}
\end{equation}

The calculation of the numerator of Eq. \eqref{eq:prob_start} is more complicated.  We again start by imposing our conditions using delta functions
\begin{align}	
	P(\mathcal{F},\text{caustic}|M) &=\int \frac{\mathcal{D}B}{\sqrt{|2\pi M|}} \ \delta(B_{3}(0)) \delta\bigg(\mathcal{F}-\frac{a_{s} \sqrt{\delta \phi} B_{\perp}^2(0)}{\sqrt{B_{3}'(0)a_{1}}}\bigg) \exp\bigg[-\frac{1}{2} B^{\dagger}M^{-1}B\bigg].
	\label{eq:prob_starting_point}
\end{align}
Using Eq. \eqref{eq:delta_function_property}, we rewrite the second delta function in terms of $B_3$
\begin{align}
	&\delta\bigg(\mathcal{F}-\frac{4 a_{s} B_{\perp}^2(0) \sqrt{\delta \phi}}{\sqrt{B_{3}'(0)a_{1}}}\bigg)=4 K \frac{B^4_{\perp}}{\mathcal{F}^3} \delta\bigg(B_{3}'(0)-K\frac{B_{\perp}^4(0)}{\mathcal{F}^2}\bigg),
\end{align}
where $K = 16 a_s^2 \delta \phi / a_1$. 

Inserting this into Eq. \eqref{eq:prob_starting_point}, we now proceed with the calculation using the Fourier representation of the delta function
\begin{align}
	&\int \frac{\mathcal{D}B}{\sqrt{|2\pi M|}} \ \delta(B_{3}(0)) \ 2 K \frac{\delta \phi^2 B^4_{\perp}}{a_1^2 \mathcal{F}^3} \ \delta\bigg(B_{3}'(0)-K\frac{B_{\perp}^4(0)}{\mathcal{F}^2}\bigg) \exp\bigg[-\frac{1}{2} B^{\dagger}M^{-1}B\bigg] \notag\\
	&=\frac{2 K \delta \phi^2}{a_1^2 \mathcal{F}^3}\int \frac{\mathcal{D}B}{\sqrt{|2\pi M|}} \int \frac{d\mu}{2\pi} \int \frac{d\eta}{2\pi} B_{\perp}^4(0) \exp\bigg[-\frac{1}{2} B^{\dagger}M^{-1}B\bigg] \exp[i\eta B_{3}(0)] \ \exp\bigg[i\mu\bigg(B_{3}'(0)-\frac{K B_{\perp}^4(0)}{\mathcal{F}^2}\bigg)\bigg]. \label{eq:step1}
\end{align}

We introduce two generalized fields $\mu'=\delta_{i3}\;\delta(x-x_{0})\;\delta(y-y_{0})\; \mu$ and $\eta'=\delta_{i3}\;\delta(x-x_{0})\;\delta(y-y_{0})\; \eta$ that permit us to work with the total magnetic field $B(z)$ rather than the LOS-component $B_{3}(z)$ in Eq. \eqref{eq:step1}: 
\begin{align}
	&\frac{4 K}{\mathcal{F}^3} \int \frac{\mathcal{D}B}{\sqrt{|2\pi M|}} \int \frac{d\mu}{2\pi} \int \frac{d\eta}{2\pi} \  B_{\perp}^4(0) \exp\bigg[-i\mu\bigg(\frac{K B_{\perp}^4(0)}{\mathcal{F}^2}\bigg)\bigg] \exp\bigg[-\frac{1}{2} B^{\dagger}M^{-1}B+i(\eta'+\mu'\partial_{3})^{\dagger}B\bigg]. \label{eq:now_we_need_perturbation_theory} 
\end{align}
The functional integral over $\mathcal{D}B$ cannot be solved analytically because of the $B_{\perp}^4(z)$-term in the exponential. In general, we need to employ diagrammatic perturbation theory in order to proceed.

If we restrict our attention to strong caustics, we can proceed analytically. This is a reasonable restriction because it will be the strong caustics that are the most likely features to be observed. We note that a strong caustic can occur when either $B_{\perp}$ is large or $B'_{3}$ is small. Strong caustics caused by an exceptionally large sky plane magnetic field are unlikely because of our assumption of Gaussian statistics, and therefore we concentrate on caustics with large $\mathcal{F}$ owing to a slowly changing LOS magnetic field.  For these caustics, we can neglect the $B_{\perp}^4(z)/\mathcal{F}^4$-term and solve the remaining problem by introducing a generating functional $J$ and repeatedly applying the Gaussian integrations
\begin{align}
	&\frac{4 K}{\mathcal{F}^3}\int \frac{\mathcal{D}B}{\sqrt{|2\pi M|}} \int \frac{d\mu}{2\pi} \int \frac{d\eta}{2\pi} \  \big(B_{1}^2(0)+B_{2}^2(0)\big)^2 \exp\bigg[-\frac{1}{2} B^{\dagger}M^{-1}B+i(\eta'+\mu'\partial_{3})^{\dagger}B\bigg] \notag\\
	&= \frac{4 K}{\mathcal{F}^3} \int \frac{\mathcal{D}B}{\sqrt{|2\pi M|}} \int \frac{d\mu}{2\pi} \int \frac{d\eta}{2\pi} \Bigg[\bigg(\frac{\delta}{\delta J_{1}(z)}\bigg)^2+ \bigg(\frac{\delta}{\delta J_{2}(z)}\Bigg)^2\bigg]^2 \exp\bigg[-\frac{1}{2} B^{\dagger}M^{-1}B+(i\eta'+i\mu'\partial_{3}+J)^{\dagger}B\bigg]\bigg|_{J=0} \notag\\
	&= \frac{4 K}{\mathcal{F}^3} \ \Bigg[\bigg(\frac{\delta}{\delta J_{1}(z)}\bigg)^2+ \bigg(\frac{\delta}{\delta J_{2}(z)}\bigg)^2\Bigg]^2 \ \int \frac{d\mu}{2\pi} \int \frac{d\eta}{2\pi} \exp\bigg[\frac{1}{2}(i\eta'+i\mu'\partial_{3}+J)^{\dagger}M(i\eta'+i\mu'\partial_{3}+J)\bigg]\bigg|_{J=0} \notag\\
	&=\frac{4 K}{\mathcal{F}^3} \ \Bigg[\bigg(\frac{\delta}{\delta J_{1}(z)}\bigg)^2+ \bigg(\frac{\delta}{\delta J_{2}(z)}\bigg)^2\Bigg]^2 \ \int \frac{d\mu}{2\pi} \int \frac{d\eta}{2\pi} \exp\bigg[-\frac{1}{2} \eta'^{\dagger}M\eta'-\frac{1}{2}(\mu'\partial_{3})^{\dagger}M(\mu'\partial_{3}) +\frac{1}{2} J^{\dagger}MJ \notag\\
	&+\underbrace{\Big(-\frac{1}{2}(\eta'+J)^{\dagger}M(\mu'\partial_{3})-\frac{1}{2}(\mu'\partial_{3})^{\dagger}M(\eta'+J)\Big)}_{=0 \text{,because} \ M_{33}(z,z') \ \text{is at its maximum for} \ z,z'=0} + i\eta'^{\dagger}MJ\bigg]\bigg|_{J=0} \notag\\
	&=\frac{4 K}{\mathcal{F}^3} \frac{1}{\sqrt{2\pi |M_{33}''(0)|}} \ \Bigg[\bigg(\frac{\delta}{\delta J_{1}(z)}\bigg)^2+ \bigg(\frac{\delta}{\delta J_{2}(z)}\bigg)^2\Bigg]^2 \int \frac{d\eta}{2\pi} \exp\bigg[-\frac{1}{2}\eta^2 M_{33}(0)+i\eta \underbrace{\int dz' J_{i} M_{i3}(z',0)}_{=I}+\frac{1}{2}J^{\dagger}MJ\bigg]\bigg|_{J=0} \notag\\
	&= \frac{4 K}{\mathcal{F}^3} \frac{1}{\sqrt{2\pi |M_{33}(0)|}} \ \frac{1}{\sqrt{2\pi |M_{33}''(0)|}} \Bigg[\bigg(\frac{\delta}{\delta J_{1}(z)}\bigg)^2+ \bigg(\frac{\delta}{\delta J_{2}(z)}\bigg)^2\Bigg]^2 \exp\bigg[\frac{1}{2}J^{\dagger}MJ-\frac{1}{2}\frac{I^2}{M_{33}(0)}\bigg]\bigg|_{J=0} \notag\\
	&= \frac{4 K}{\mathcal{F}^3} \frac{1}{\sqrt{2\pi |M_{33}(0)|}} \ \frac{1}{\sqrt{2\pi |M_{33}''(0)|}} \ \Big[3\big(M_{11}^2(0)+M_{22}^2(0)\big) + M_{22}(0)M_{11}(0)+\frac{1}{M_{33}^2(0)}\underbrace{\big(M_{13}^4(0)+M_{23}^4(0)+2M_{23}^2(0)M_{13}^2(0)\big)}_{=0}\Big] \notag\\
	&=\frac{28 K}{\mathcal{F}^3} \frac{1}{\sqrt{2\pi |M_{33}(0)|}} \ \frac{1}{\sqrt{2\pi |M_{33}''(0)|}} M_{N}^2 = \frac{28 K}{\mathcal{F}^3} \frac{1}{\sqrt{2\pi |M_{33}(0)|}} \ \frac{1}{\sqrt{2\pi |M_{33}''(0)|}} \left[ \int \frac{dk^3}{(2\pi)^3} \ \hat{M}_{N}(k) \right]^2 \notag\\
	&=\frac{28 K}{\mathcal{F}^3} \frac{1}{\sqrt{2\pi |M_{33}(0)|}} \ \frac{1}{\sqrt{2\pi |M_{33}''(0)|}} \left[ \int dk^3 \ \frac{\epsilon_{B}(k)}{k^2} \right]^2.
\end{align}
In the last step, we identified the covariance matrix $M$ with the magnetic correlation tensor given by Eqs. \eqref{eq:magnetic_correlation_tensor} and \eqref{eq:eb_to_MN}.  We used the property $M_{ij}(0)=M_{N}(0) \ \delta_{ij}$. 
Finally, from this result and Eq. \eqref{eq:prob_spike}, we conclude that in the bright caustic limit

\begin{align}
	P\left(\mathcal{F}|\text{caustic}, M\right) &= \frac{P\left(\mathcal{F}, \text{caustic}|M\right)}{P\left(\text{caustic}|M\right)} = \frac{448 a_s^2 \delta \phi^2}{\mathcal{F}^3 a_1} \frac{M^2_N(0)}{\sqrt{2\pi|M''_{33}(0)|}} \notag\\
	&= \frac{448 a_s^2 \delta \phi^3}{\mathcal{F}^3 a_1^2\sqrt{2\pi|M''_{33}(0)|}} \left[\int dk^3 \frac{\epsilon_{B}(k)}{k^2}\right]^2 = \frac{448 a_s^2 \delta \phi^3}{\mathcal{F}^3 a_1^2\sqrt{2\pi}}\frac{\left[\int dk^3 \frac{\epsilon_{B}(k)}{k^2}\right]^2}{\left[\int dk^3 \epsilon_{B}(k)\right]^{1/2}}.
	\label{eq:prob_result}
\end{align}

\clearpage
\section{Simulated Faraday spectral cube slices}\label{sec:cube_slices}

\begin{figure}[h!]
	\centering
	\subfigure{\includegraphics[width=15cm]{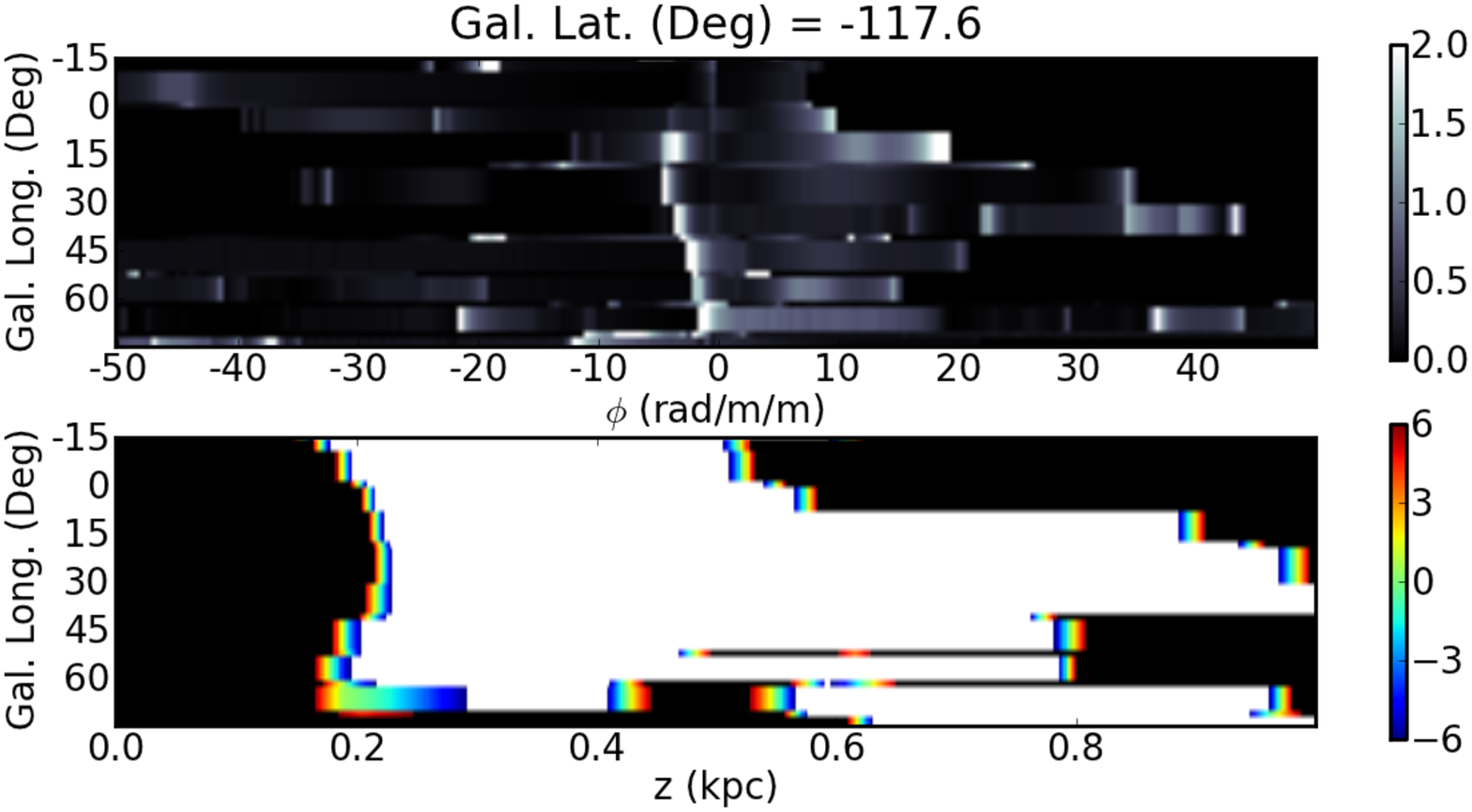}}
	\subfigure{\includegraphics[width=15cm]{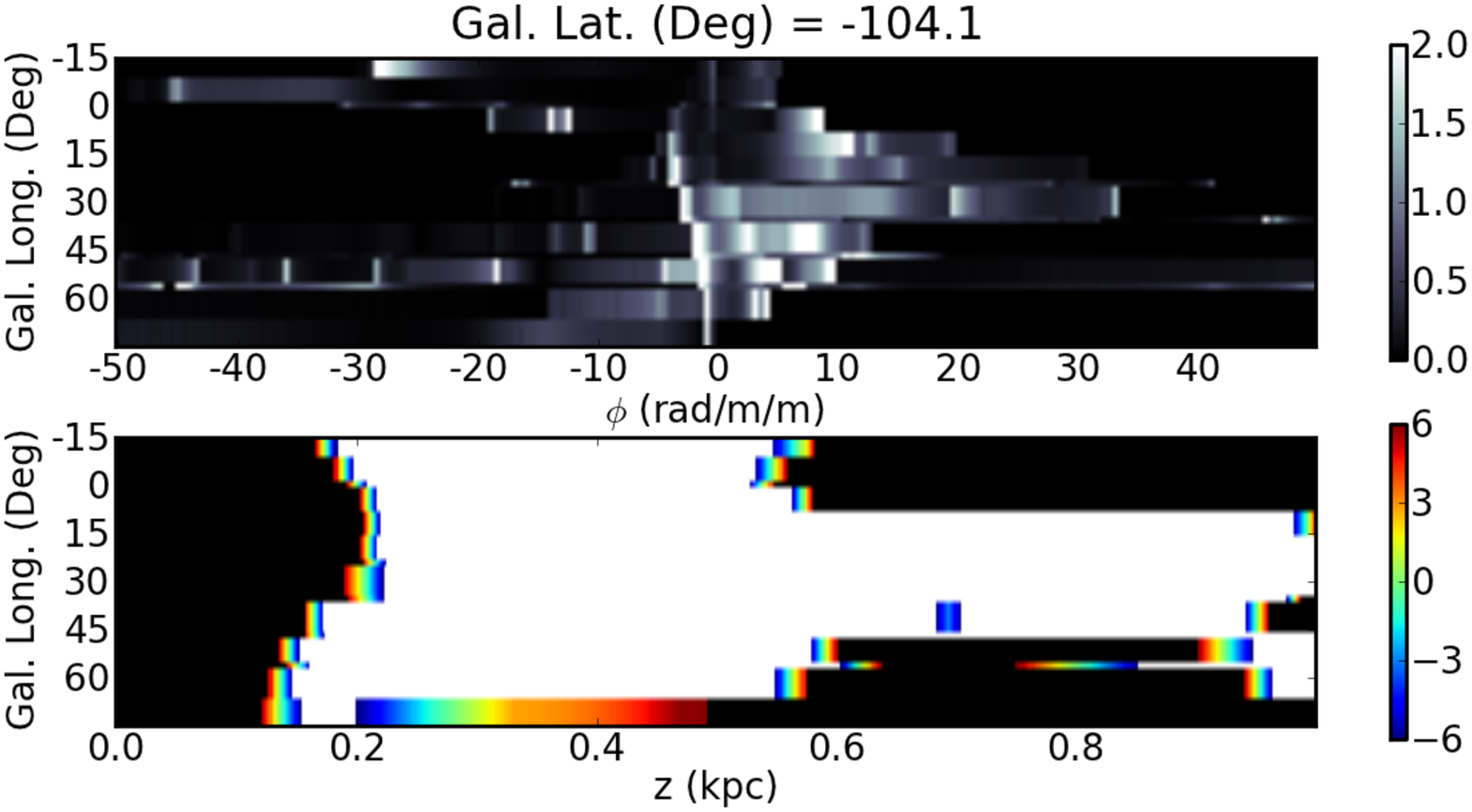}}
	\subfigure{\includegraphics[width=15cm]{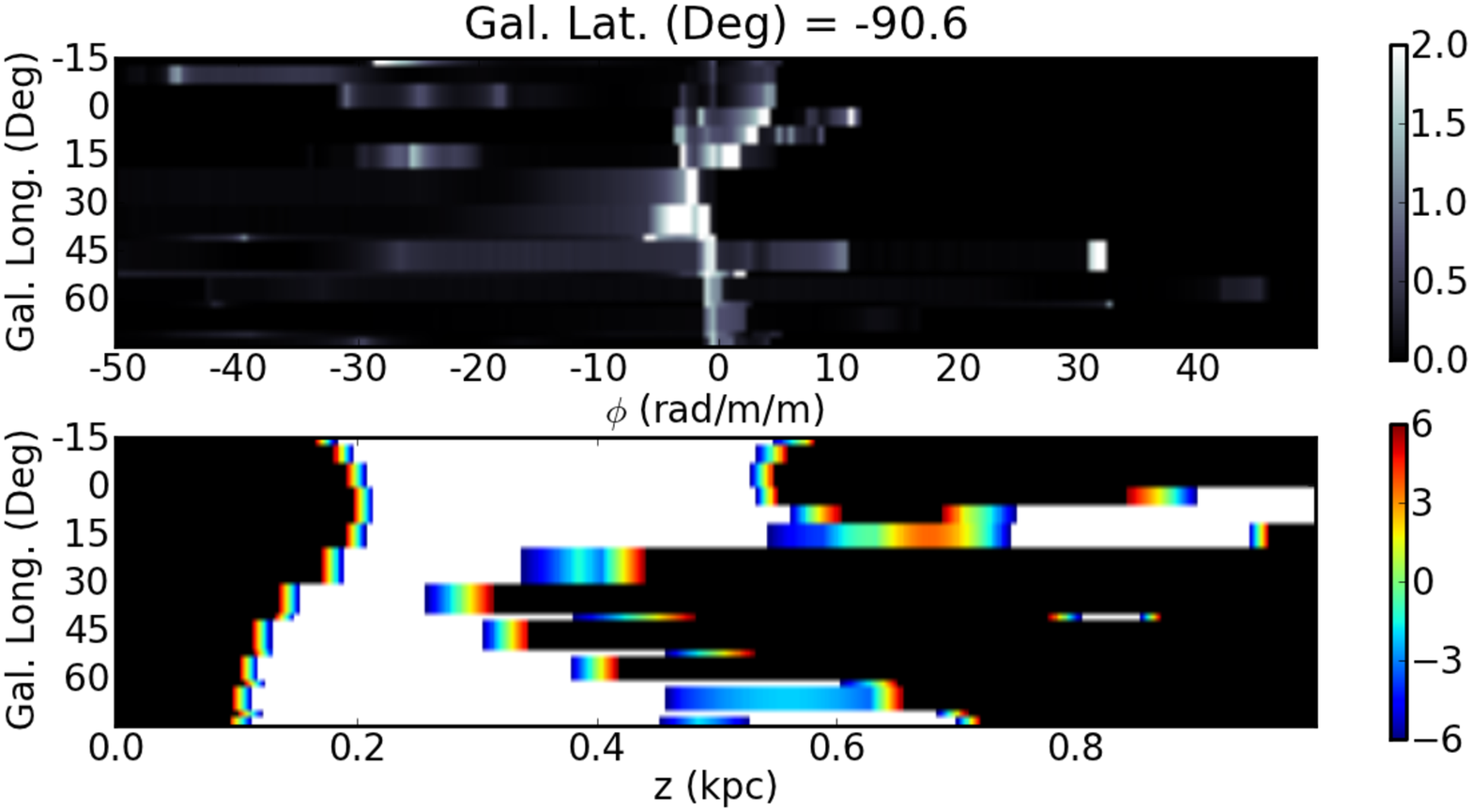}}
	\caption{The first three slices of a Faraday spectrum cube of a Hammurabi simulation and the associated LOS magnetic field strength.  Each pair shows a frame at a different Galactic latitude.  Top: A grey scale image of polarized intensity (in arbitrary units) as a function of Faraday depth and Galactic longitude.  Bottom: The LOS magnetic field strength as a function of LOS distance (in pc) and Galactic longitude.  The scale has been compressed to highlight regions with opposite polarity.  Black indicates positive magnetic field strengths and white negative.  The colored area is where $B_3 \approx 0$ (appears as grey scale in text version).}
	\label{fig:movie_slices1}
\end{figure}

\clearpage
\begin{figure}[h!]
	\centering
	\subfigure{\includegraphics[width=15cm]{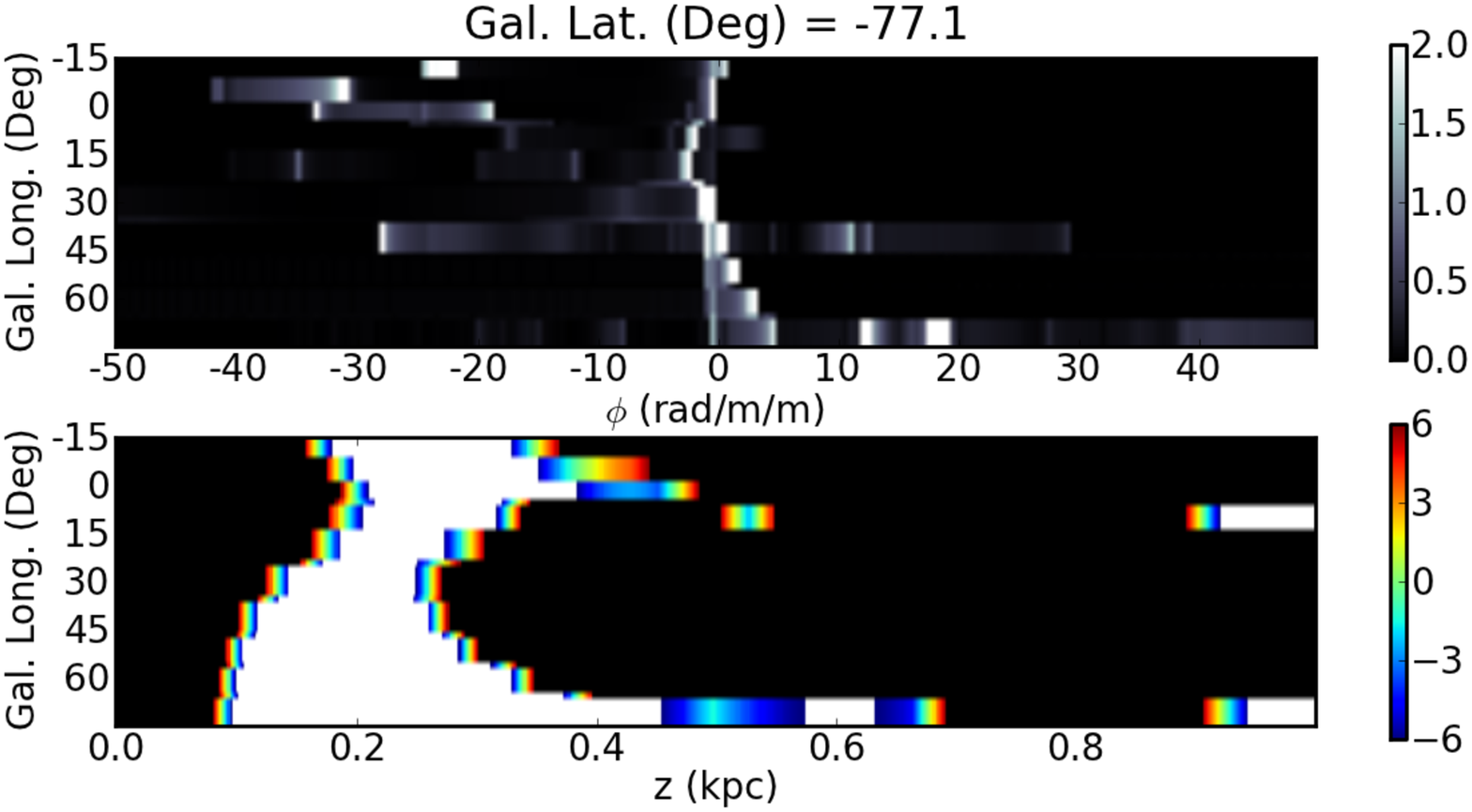}}
	\subfigure{\includegraphics[width=15cm]{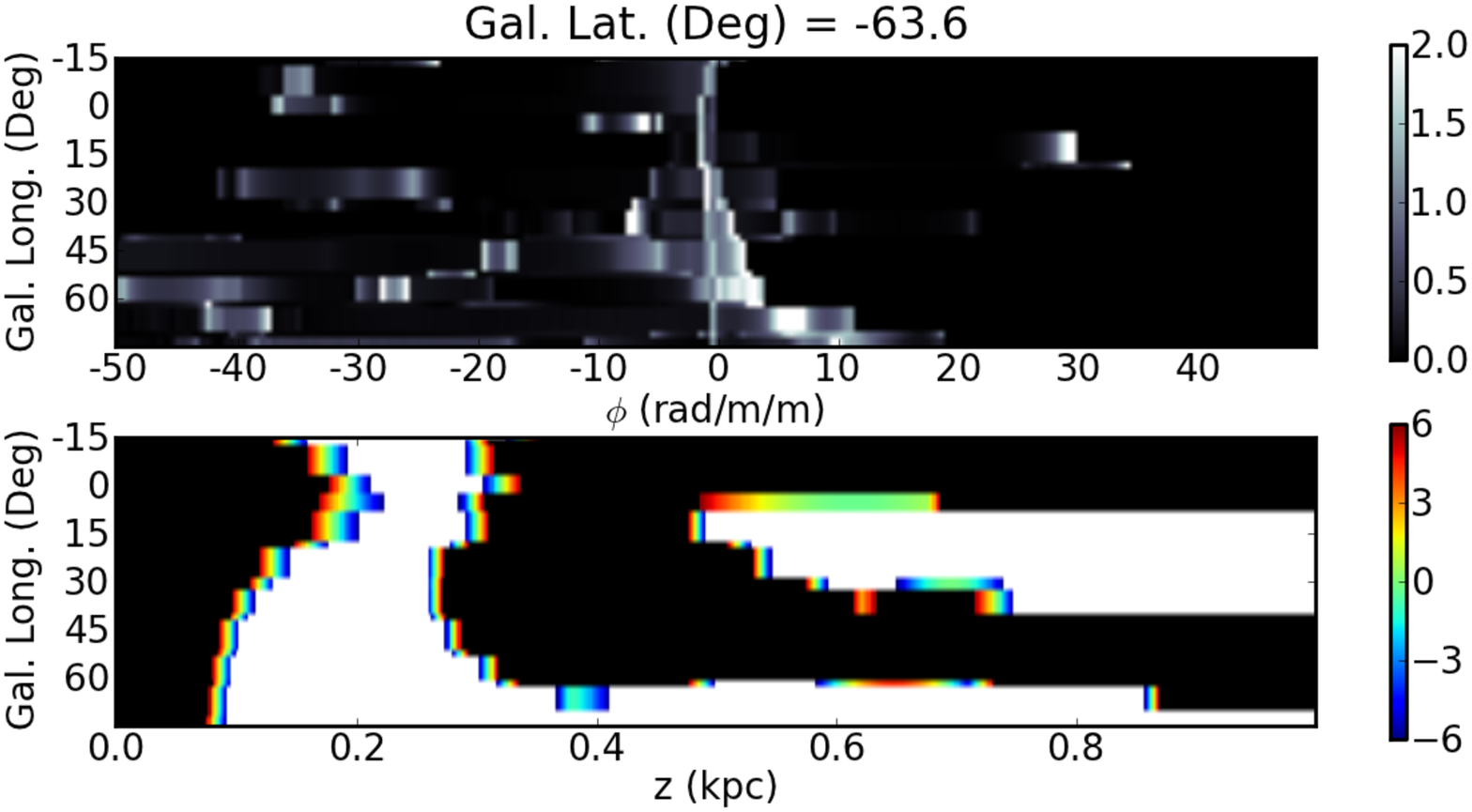}}
	\subfigure{\includegraphics[width=15cm]{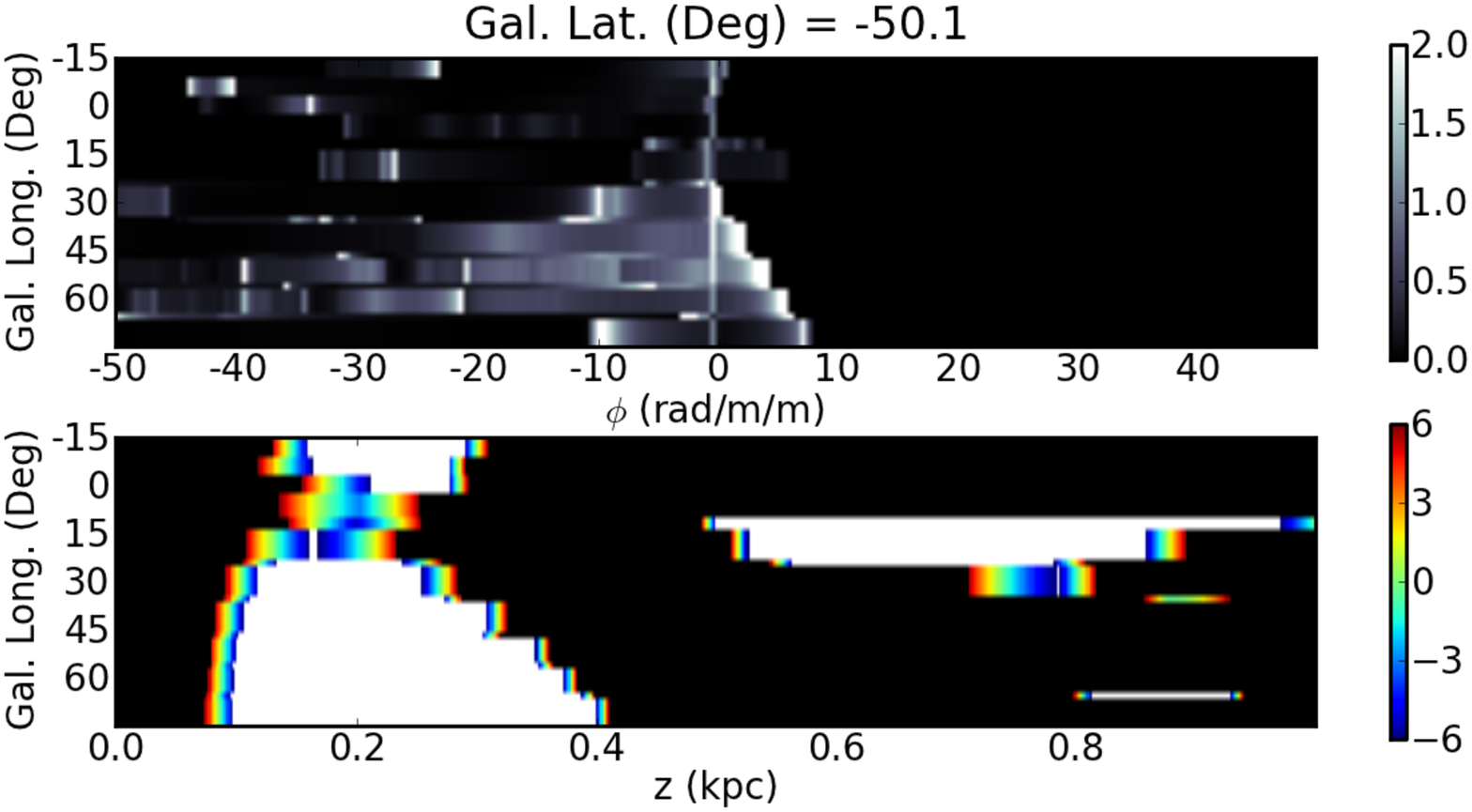}}
	\caption{Three more slices of a Faraday spectrum cube of a Hammurabi simulation, following those in Fig. \ref{fig:movie_slices1}.}  
	\label{fig:movie_slices2}
\end{figure}

\end{document}